\documentclass[11pt,english]{article}
\usepackage{amsfonts}
\usepackage{float}
\usepackage{amsmath}
\usepackage{amssymb}
\usepackage{graphicx}
\usepackage{setspace}
\usepackage[authoryear]{natbib}
\usepackage{appendix}
\usepackage{grffile}
\usepackage{natbib}
\usepackage{thmtools}

\newtheorem{proposition}{Proposition}
\newtheorem{example}{Example}

\allowdisplaybreaks
\input{tcilatex}
\begin{document}

\title{Moment conditions and Bayesian nonparametrics\thanks{%
We thank Isaiah Andrews, Yang Chen, Herman van Dijk, Mikkel Plagborg-Moller
and Christian Robert for their comments on an earlier draft. \ }}
\author{\textsc{Luke Bornn} \\
\textit{Department of Statistics, Harvard University and } \\
\textit{Department of Statistics and Actuarial Science, Simon Fraser
University }\\
\texttt{bornn@fas.harvard.edu} \and \textsc{Neil Shephard} \\
\textit{Department of Economics and Department of Statistics, Harvard
University}\\
\texttt{shephard@fas.harvard.edu} \and \textsc{Reza Solgi} \\
\textit{Department of Statistics, Harvard University}\\
\texttt{rezasolgi@fas.harvard.edu}}
\maketitle

\begin{abstract}
Models phrased though moment conditions are central to much of modern
inference. Here these moment conditions are embedded within a nonparametric
Bayesian setup. Handling such a model is not probabilistically
straightforward as the posterior has support on a manifold. We solve the
relevant issues, building new probability and computational tools using
Hausdorff measures to analyze them on real and simulated data. These new
methods which involve simulating on a manifold can be applied widely,
including providing Bayesian analysis of quasi-likelihoods, linear and
nonlinear regression, missing data and hierarchical models.
\end{abstract}

\renewcommand\thmcontinues[1]{Continued}


\baselineskip=20pt

\section{Introduction}

\subsection{Overview}

Much of modern inference is phrased in terms of moment conditions and
analyzed using asymptotic approximations. Here we build a new methodology
which dovetails with decision theory. Moment conditions are embedded within
a nonparametric Bayesian setup, allowing an individual to mix moment
conditions with data and scientifically informative priors to make rational
decisions without the recourse to the veil of parametric assumptions or
asymptotics. \ 

Embedding moments within nonparametrics is not probabilistically
straightforward. This paper spells out the issues, develops the
corresponding probability theory to solve them and devises novel strategies
for simulating on a manifold to implement them in practice on simulated and
real data. \ It covers the case where it is hard, or indeed impossible, to
solve the moment equations. This allows the rational analysis of moment
condition models with many solutions.

The scope of the new methods is vast. It deals with, for example, linear,
nonlinear and instrumental variable regression. By thinking of the moment
condition as the score of a parametric statistical model, our analysis also
provides a Bayesian treatment of quasi-likelihood methods which are widely
applied in statistics (e.g. \cite{Cox(61)}, \cite{White(94)}). Finally, this
framework provides a solid basis to deal systematically with missing data
(e.g. \cite{LittleRubin(02)}), shrink parameters (e.g. \cite{Efron(12)}) and
build hierarchical models (e.g. \cite%
{GelmanCarlinSternDunsonVehtariRubin(13)}). \ 

\subsection{The conceptual challenge\ }

It will be helpful in our discussion of the paper's contribution and to
place it in the context of the literature to establish some notation; a
formal statement will appear in Section \ref{sect:Bayes moment}. \ 

Assume one has independent and identically distributed (i.i.d.) $d$%
-dimensional data $Z_{i}$, $i=1,2,...,n$, taking on the known support $%
s_{1},s_{2},...,s_{J}$ and having distribution function $F$. We then write $%
\mathbb{P}(Z_{i}=s_{j}|\theta ,\beta )=\theta _{j}$ where the $p$%
-dimensional $\beta $ satisfies the $r$-dimensional moment condition 
\begin{equation}
\mathbb{E}_{Z}\left\{ g(Z,\beta )\right\} =\int g(z,\beta )F(\mathrm{d}%
z)=\sum_{j=1}^{J}\theta _{j}g(s_{j},\beta )=0.  \label{moment condition}
\end{equation}%
Here $\beta $ is the parameter of scientific interest. We then view $\theta
=(\theta _{1},\theta _{2},...,\theta _{J-1})^{\prime }$ (with $\theta
_{J}=1-\iota ^{\prime }\theta $, where $\iota $ is a vector of ones of
appropriate size) as nuisance parameters to be treated nonparametrically.
The task is to learn $p(\beta ,\theta |Z)$ or $p(\beta |Z)$, where $%
Z=(Z_{1},Z_{2},...,Z_{n})^{\prime }$. A simple example of this is $%
g(s_{j},\beta )=s_{j}-\beta $ which delivers the mean. \ 

Although this problem is easy to state, it is not easily carried through, as
traditional nonparametric models clash with the moment conditions, in effect
overspecifying the model. Expressing this in a different way: the prior and
posterior for $\beta ,\theta $ are typically supported on a zero Lebesgue
measure $(J+p-1-r)$-dimensional set, $\Theta _{\beta ,\theta }$, in $\mathbb{%
R}^{J+p-1}$. As a result, traditional Markov chain Monte Carlo (MCMC)
methods (or alternatives like importance sampling) for sampling from $%
p(\beta ,\theta |Z)$ entirely collapse. This paper solves this problem in
two different ways: the comparative advantages of each will depend upon the
form of the moment conditions. Taken overall this paper provides a unified
solution to this central problem. \ 

\subsection{Literature on classical analysis of moments}

Before we detail our new approach, we will discuss how this work relates to
the literature. \ 

Moment based estimation was introduced by \cite{Pearson(1894)}. A relatively
modern version of this procedure first estimates $\widehat{\theta }$
nonparametrically, that is $F$ by the empirical distribution function $F_{n}$%
, and then plugs it into (\ref{moment condition}), yielding the function 
\begin{equation*}
\int g(z,\beta )F_{n}(\mathrm{d}z)=\sum_{j=1}^{J}\widehat{\theta }%
_{j}g(s_{j},\beta ).
\end{equation*}%
In the $p=r$ case we move $\beta $ around until this function equals a
vector of zeros, delivering the method of moments estimator $\widehat{\beta }
$. Extensions include, for example, \cite{Sargan(58),Sargan(59)}, \cite%
{Durbin(60)}, \cite{Godambe(60)}, \cite{Wedderburn(74)}, \cite%
{McCullaghNelder(89)}, \cite{Hansen(82)}, \cite{Chamberlain(87)}, \cite%
{HansenHeatonYaron(96)}, \cite{GallantTauchen(96)} and \cite%
{GourierouxMonfortRenault(93)}. \cite{Hall(05)} gives a recent review. \ 

An elegant implementation of moment based inference is through empirical
likelihood. Motivated by \cite{Owen(88),Owen(90)}, \cite{QinLawless(94)} and 
\cite{ImbensSpadyJohnson(98)} discussed empirical likelihood based inference
in overidentified moment condition models. See also the reviews by \cite%
{Owen(01)}, \cite{Kitamura(07)} and \cite{LancasterJun(10)}.

\subsection{Literature on Bayesian analysis of moments}

Our work is fully Bayesian. Much of our work has been inspired by \cite%
{Chamberlain(87)} and in particular \cite{ChamberlainImbens(03)}. \cite%
{ChamberlainImbens(03)} place a Dirichlet prior on $\theta $, which implies
the posterior on $\theta $ is Dirichlet. These priors and posteriors are
straightforward to sample from as noticed by \cite{Rubin(81)} in his
Bayesian bootstrap. \cite{ChamberlainImbens(03)} suggest that for each
posterior draw of $\theta $ they would solve the moment conditions to imply
a value (or in principle a set of values) of $\beta $. Collecting a sample
of such solved values provides a sample from a posterior on $\beta $.
Unfortunately these authors have no control over the prior for $\beta $, the
parameter of scientific interest.

Also important is \cite{KitamuraOtsu(11)}, who have two methods, both
expressed in terms of Dirichlet process priors. Here we convert them into
our finite framework. In their exponentially tilted case they first specify
a prior $p(\beta )p(\theta )$ before finding $\theta ^{\ast }=\left( \theta
_{1}^{\ast },\theta _{2}^{\ast },...,\theta _{J}^{\ast }\right) $ which
minimizes $\sum_{j=1}^{J}\theta _{j}^{\ast }\log \left( \frac{\theta
_{j}^{\ast }}{\theta _{j}}\right) $ subject to the moment constraints $%
\sum_{j=1}^{J}\theta _{j}^{\ast }g(s_{j},\beta )=0$ and the probability
axioms. They then set $\mathbb{P}(Z_{i}=s_{j}|\theta ,\beta )=\theta
_{j}^{\ast }$, using this model to learn $\beta $ and $\theta $ from the
data. \cite{Shin(14)} carefully investigates various computational aspects
of this approach. This approach has many advantages but it leaves pairs of $%
\beta $ and $\theta $ with positive posterior probability which are not
logically compatible. \cite{KitamuraOtsu(11)} also propose a synthetic
Dirichlet process (with connections to \cite{Doss(85)} and \cite%
{NewtonCzadoChappell(96)}).

There are also many papers which provide alternative methods, including a
substantial literature on the Bayesian use of moments through approximate
methods. \cite{ChernozhukovHong(03)} specify a quadratic form in the moment
conditions and use this as the basis of a log quasi-likelihood function.
They then use this approximate likelihood to carry out Bayesian inference
using MCMC alongside a sandwich estimator. Related work includes \cite%
{Yin(09)}. \cite{Muller(13)} provides a Bayesian version of the asymptotic
sandwich matrix commonly seen in quasi-likelihood inference and links it to
decision theory. \ 

\cite{Lazar(03)}, \cite{Schennach(05)} and \cite{YangHe(12)} provide
Bayesian interpretations to empirical likelihood and study the resulting
properties. \cite{MengersenPudloRobert(13)} look at moment conditions and
empirical likelihood using approximate Bayesian computation. See also \cite%
{Zellner(97)} and \cite{ZellnerTobiasRyu(97)}, who suggested a Bayesian
moment method by building a likelihood defined through the maximum entropy
density consistent with the moment conditions. \ Related is the Bayesian
work on factor and cointegration models, e.g. \cite{StrachanvanDijk(04)}.

In a series of papers \cite{GallantHong(07)}, \cite%
{GallantGiacominiRagusa(14)} and \cite{Gallant(15)} develop methods which
devise a prior using fiducial arguments from moment conditions. Related work
includes \cite{Jaynes(03)} and \cite{Kwan(98)}. \cite{FlorenSimoni(15)} have
used Gaussian processes in combination with moment constraints to carry out
Bayesian inference.

\subsection{Computational issues}

Here the prior and posterior for $\beta ,\theta $ are supported on a zero
Lebesgue measure $(J+p-1-r)$-dimensional set, $\Theta _{\beta ,\theta }$, in 
$\mathbb{R}^{J+p-1}$. Hence Bayesian inference will need us to sample from a
distribution defined on a zero measure set, rendering standard Monte Carlo
methods useless. \ \ \ 

In an influential paper \cite{GelfandSmithLee(92)} use MCMC methods to deal
with constrained parameter spaces, but in their paper the constraints do not
change the dimension of the support. \cite{HurnRueSheehan(99)} carry out
MCMC in constrained parameter spaces (sampling from a distribution $\pi (x)$
subject to a constraint $C(x)=0$) using block updating. \ \cite%
{GolchiCampbell(14)} carry out sampling subject to constraints using
sequential Monte Carlo methods by slowly introducing the constraints.
However, they do not explore the change of measure issue we discuss here. 
\cite{Chiu(08)} use a singular normal distribution in posterior updating for
an under-identified hierarchical model. Related work includes \cite%
{SunSpeckmanTsutakawa(99)}. Overspecified factor models also have some of
these features, as discussed by \cite{West(03)}.\ {\cite%
{FiorentiniSentanaShephard(04)}} face related but highly specialized
challenges when sampling missing data in a GARCH model. \ 

There are few recent papers on MCMC simulation from distributions defined on
manifolds. \cite{BrubakerSalzmannUrtasun(12)} propose a Hamiltonian Monte
Carlo on implicitly defined manifolds. Numeric integration of the
Hamiltonian dynamics requires solving a system of $3d$ nonlinear equations
for each update, where $d$ is the dimension of the space in which the
manifold is embedded (in our setting $d=J+p-1$). \cite{ByrneGirolami(13)}
introduce a Hamiltonian Monte Carlo simulation algorithm for sampling from
manifolds with known geodesic structure. They demonstrate how this algorithm
can be used in order to sample from the distributions defined on
hyperspheres and Stiefel manifolds of orthonormal matrices. \cite%
{DiaconisHolmesShahshahani(13)} provide a short review of concepts in
geometric measure theory. They discuss algorithms for sampling from
distributions defined on Riemannian manifolds that are similar to the
\textquotedblleft marginal method" that will be introduced shortly.\ It is
this paper which has been the most helpful to us in terms of Monte Carlo
methods. \ 

\subsection{Outline of the paper}

In the next section of the paper we will introduce the formal model under
study, and discuss how one specifies meaningful prior distributions on the
parameters of interest. In Section \ref{sect:inference} several methods for
inference and their relative merits and pitfalls are discussed. Section \ref%
{sect:priors} discusses mechanisms for generating priors for these models.
We also draw out how to make inference when the support of the data is
unknown, regarding the unseen support as missing. This is followed by
Section \ref{sectio:examples} in which some illustrative examples are
demonstrated. Section \ref{sectio: empirical studies} explores several
empirical studies before Section \ref{section:conclusions} concludes. An
Appendix collects the proofs of the propositions stated in the paper and a
collection of additional results. \ 

\section{Bayesian moment conditions models\label{sect:Bayes moment}}

\subsection{The model}

Assume the data we have available to make inference is $Z=\left(
Z_{1},...,Z_{n}\right) $, where the $Z_{i}$ are $d$-dimensional i.i.d. draws
from an unknown distribution which has $J$ points of known support $%
\{s_{1},s_{2},...,s_{J}\}=S$ (we relax this known support condition in
Section \ref{sect:missing support}). Throughout we write 
\begin{equation}
\mathbb{P}(Z_{i}=s_{j}|\theta ,\beta )=\theta _{j},\quad j=1,2,...,J,
\end{equation}%
with $\theta =(\theta _{1},\theta _{2},...,\theta _{J-1})^{\prime }\in
\Theta _{\theta }\subseteq \Delta ^{J-1}$, where $\Delta ^{J-1}=$\{$\theta
=(\theta _{1},\theta _{2},...,\theta _{J-1})^{\prime }$;$\ \iota ^{\prime
}\theta <1\ $and$\ \theta _{j}>0$\}$\ $for all $j$ and $\theta _{J}=1-\iota
^{\prime }\theta $, in which $\iota $ is a vector of ones. Further, the
science of the problem is characterized by the values of $\beta $ which
solve the $r$ unconditional moment conditions, 
\begin{equation}
\sum_{j=1}^{J}\theta _{j}g(s_{j},\beta )=0,
\end{equation}%
where $\beta \in \Theta _{\beta }\subseteq \mathbb{R}^{p}$ and $g:\mathbb{R}%
^{d}\times \mathbb{R}^{p}\rightarrow \mathbb{R}^{r}$. Typically the
scientific conclusions will center around inferences on $\beta $, although
predictive type inference may also additionally feature $\theta $.\ This
paper concentrates on the case of exactly identified models ($r=p$).
Appendix \ref{sect:not just identified} extends to the more general case of
over and under identification at the cost of more clutter but without having
to generate any new ideas. \ 

\subsection{Parameter space and prior}

Throughout this paper we will think of $\beta $ and $\theta $ as parameters
to be learned from the data, $Z$. \ We write the $J+p-1$ parameters 
\begin{equation*}
(\beta ^{\prime },\theta ^{\prime })^{\prime }\in \Theta _{\beta ,\theta },
\end{equation*}%
where $\Theta _{\beta ,\theta }\subseteq \mathbb{R}^{p}\times \Delta
^{J-1}\subset \mathbb{R}^{J+p-1}$, as the joint support for $\beta $ and $%
\theta $. Each point within $\Theta _{\beta ,\theta }$ is a pair $(\beta
,\theta )$ which satisfies both the moment conditions and probability
axioms. The moment conditions are: 
\begin{equation*}
H_{\beta }\theta +g_{J}=0\quad \text{where\quad }H_{\beta }=\left(
g_{1},...,g_{J-1}\right) -g_{J}\iota ^{\prime },
\end{equation*}%
in which $g_{j}=g(s_{j},\beta )$ (for $1\leq j\leq J$). 
Moreover $H_{\beta }$ is assumed to be of full row rank (we will often
suppress the dependence on $\beta $ and just write $H$). These constraints,
together with the inequalities $\theta _{j}\geq 0$ (for $j=1,2,...,J$),
implicitly define the $(J-1)$-dimensional set of parameters within $\mathbb{R%
}^{J+p-1}$, which will be denoted by $\Theta _{\beta ,\theta }$. Hence the
parameter space, $\Theta _{\beta ,\theta }$, depends upon the support of the
data, $S=\{s_{1},...,s_{J}\}$, but is not data dependent. Throughout the
paper, the notation $\Theta _{\lambda }$ will generically represent the
parameter space of $\lambda $ in which $\lambda $ is a set of parameters.

The set of admissible pairs $(\beta ,\theta )$, denoted by $\Theta _{\beta
,\theta }$, is a zero measure set (with respect to Lebesgue measure) in $%
\mathbb{R}^{J+p-1}$. We will assume that researchers can place a prior
density, $p(\beta ,\theta )$, with respect to the $(J-1)-$dimensional
Hausdorff measure on $\Theta _{\beta ,\theta }$. Using the Hausdorff measure%
\footnote{%
Assume $E\subseteq \mathbb{R}^{n}$, $d\in \lbrack 0,+\infty )$ and $\delta
\in (0,+\infty ]$. The Hausdorff premeasure of $E$ is defined as follows, 
\begin{equation*}
\mathcal{H}_{\delta }^{d}(E)=v_{m}\ \inf_{\substack{ E\subseteq \cup E_{j} 
\\ d(E_{j})<\delta }}\ \ \sum_{j=1}^{\infty }\left( \frac{\text{diam}(E_{j})%
}{2}\right) ^{d}
\end{equation*}%
where $v_{m}=\frac{\Gamma (\frac{1}{2})^{d}}{2^{d}\Gamma (\frac{d}{2}+1)}$
is the volume of the unit $d$-sphere, and $\text{diam}(E_{j})$ is the
diameter of $E_{j}$. $\mathcal{H}_{\delta }^{d}(E)$ is a nonincreasing
function of $\delta $, and the $d$-dimensional Hausdorff measure of $E$ is
defined as its limit when $\delta $ goes to zero, $\mathcal{H}^{d}(E)=\lim 
_{\substack{ \delta \rightarrow 0^{+}}}\ \mathcal{H}_{\delta }^{d}(E)$. The
Hausdorff measure is an outer measure. Moreover $\mathcal{H}^{n}$ defined on 
$\mathbb{R}^{n}$ coincide with Lebesgue measure. See \cite{Federer(69)} for
more details. \ } as the base measure, we are able to assign measures to the
lower dimensional subsets of $\mathbb{R}^{^{J+p-1}}$, and therefore we can
define probability density functions with respect to Hausdorff measure on
manifolds (and more complex zero Lebesgue measure sets) in an Euclidean
space.

\subsection{Some examples}

To cement this we have built a starkly simple example which captures most of
the challenges in this problem. It faces off a nonparametric model against a
scientific parameter of interest. \ \ 

\begin{example}
\label{exa:ExampleLogit1} \textbf{(Logistic)} Assume $Z_{1}|\theta \sim 
\mathrm{Bernoulli}(\theta )$, and let $\beta =\log \left( \frac{\theta }{%
1-\theta }\right) =\mathrm{logit}(\theta )$ be the scientific parameter of
interest. Jointly $\beta ,\theta $ captures the inherent singularity
implicit in all moment based inference. The moment condition is 
\begin{equation*}
g(s,\beta )=s-\frac{e^{\beta }}{1+e^{\beta }}.
\end{equation*}%
Therefore the parameter space, $\Theta _{\beta ,\theta }$, is 
\begin{equation*}
\Theta _{\beta ,\theta }=\left\{ (\beta ,\theta )\in \mathbb{R}\times
\lbrack 0,1];\beta =\log \left( \frac{\theta }{1-\theta }\right) \right\} .
\end{equation*}%
This is shown as the blue curve sitting at ground level in the left panel of
Figure \ref{Fig:support_density}. \ 
\begin{figure}[tbp]
\begin{center}
\includegraphics[width=8cm]{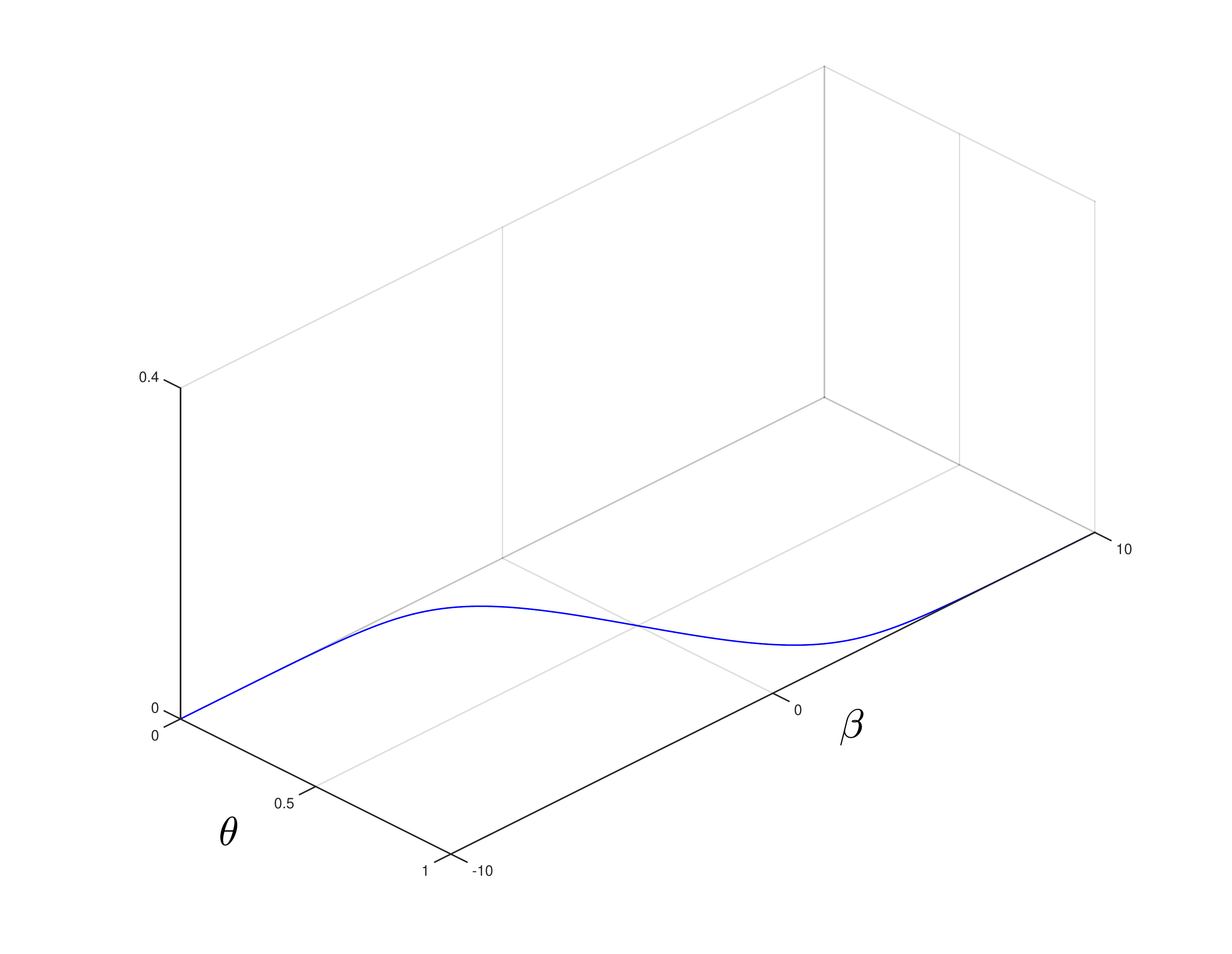} %
\includegraphics[width=8cm]{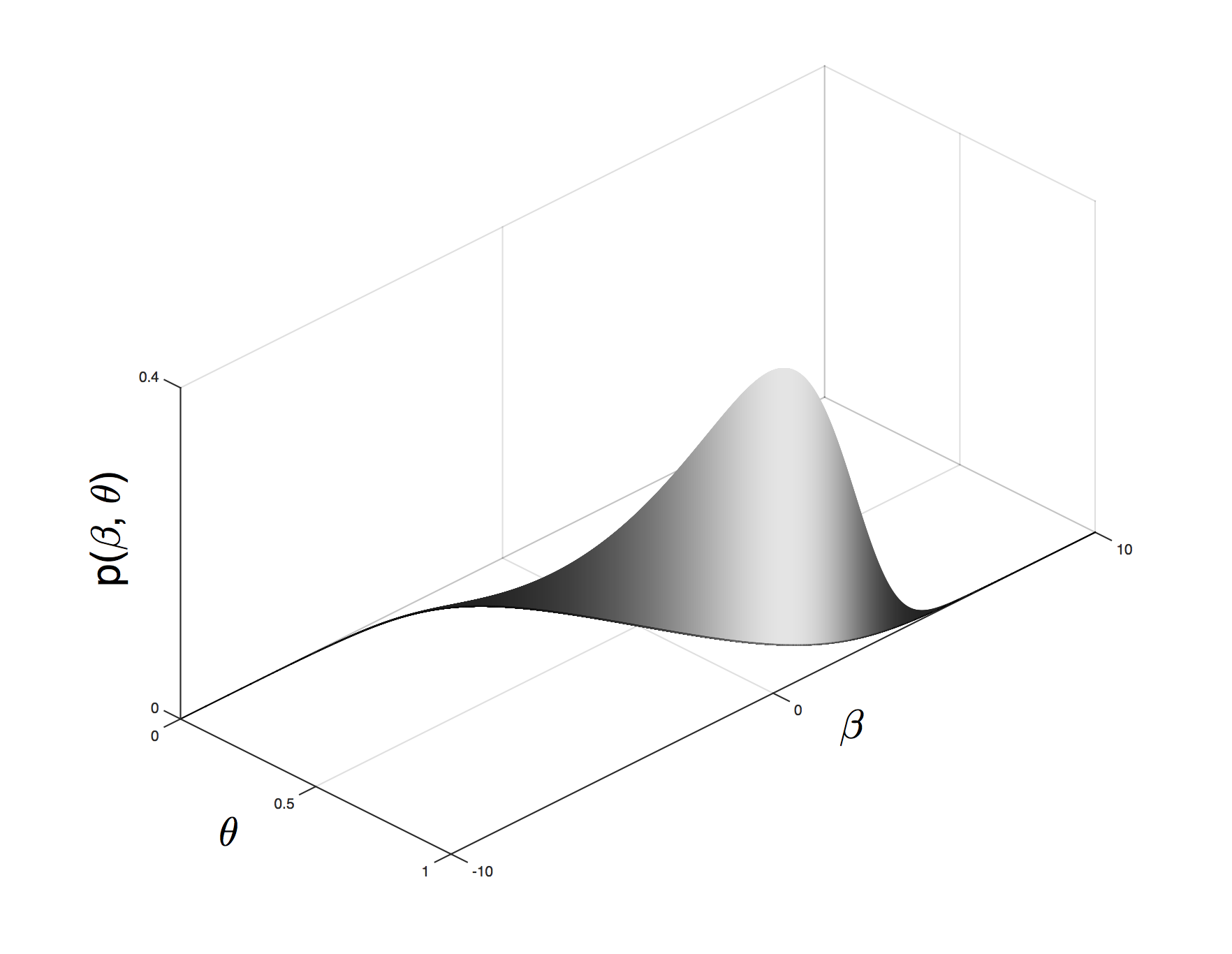}
\end{center}
\caption{In the plot on the left the blue curve, $\protect\beta =\log \left( 
\frac{\protect\theta }{1-\protect\theta }\right) $, is the parameter space
of the logit model, $\Theta _{\protect\beta ,\protect\theta }$. In the plot
on the right the density of the prior $p(\protect\beta ,\protect\theta )$
(with respect to Hausdorff measure) is depicted. This density lives on the
blue curve which supports $\Theta _{\protect\beta ,\protect\theta }$.}
\label{Fig:support_density}
\end{figure}
Of fundamental importance is that if $\theta $ moves by $\mathrm{d}\theta $
then the length of the journey along this curve will be (by Pythagoras's
theorem) 
\begin{equation*}
\mathrm{d}\theta \sqrt{1+J_{\theta }^{2}},\quad J_{\theta }=\frac{\partial
\beta }{\partial \theta }=\frac{\partial \log \left( \frac{\theta }{1-\theta 
}\right) }{\partial \theta }.
\end{equation*}%
The right panel of Figure \ref{Fig:support_density} repeats the support but
now above it is a (the form of the density is not expositionally important
at this point) density $p(\beta ,\theta )$ with respect to this curve, or
more formally the one dimension Hausdorff measure on $\Theta _{\beta ,\theta
}$. \ 
Then for any set $C\subset \Theta _{\beta ,\theta }$, 
\begin{equation*}
\Pr \left\{ \left( \beta ,\theta \right) \in C\right\} =\int_{C_{\theta
}}p(\beta ,\theta )\sqrt{1+\left( \frac{\partial \beta }{\partial \theta }%
\right) ^{2}}\mathrm{d}\theta ,
\end{equation*}%
where $C_{\theta }$ is the projection of $C$ on $\theta $'s axis (i.e. we
integrate over all values of $\theta $ which imply a $\beta $ such that the
pair $\left( \beta ,\theta \right) \in C$). This means as we integrate over $%
\theta $, we must multiply the density on the curve by the length of the
curve. \ 
\end{example}

We will study how to transform this prior $p(\beta ,\theta )$ into a
posterior and simulate from it. This will allow us to learn $\beta $ from
the data. As with all Bayesian calculations, it is not trivial to establish
a widely acceptable prior $p(\beta ,\theta )$. We will return to that very
practical issue in Section \ref{sect:priors}. \ 

Before we leave this section we give a less artful example.

\begin{example}
\label{exa:ExampleMeanJ3} \textbf{(Mean)} Let $Z$ be a scalar random
variable and $g(s,\beta )=s-\beta $, so $\beta $ is a mean. Then 
\begin{equation*}
\Theta _{\beta ,\theta }=\left\{ (\beta ,\theta )\ ;\ \sum_{j=1}^{J}\theta
_{j}s_{j}=\beta ,\ \theta _{j}>0\ \text{ for all }j,\text{ and }\iota
^{\prime }\theta <1\ \right\} .
\end{equation*}%
Thus $\Theta _{\beta ,\theta }$ is a region within a $(J-1)$-dimensional
hyperplane in $\mathbb{R}^{J}$. However all elements of this set are not
admissible, since $\theta $ should satisfy the probability axioms (elements
of $\theta $ should be positive and $1-\iota ^{\prime }\theta >0$).
Therefore the parameter space $\Theta _{\beta ,\theta }$ is a convex subset
on the hyperplane. Then if $\theta $ moves by $\mathrm{d}\theta _{1},...,%
\mathrm{d}\theta _{J-1}$ the area of the corresponding parallelogram on the
hyperplane is 
\begin{equation*}
\mathrm{d}\theta _{1}...\mathrm{d}\theta _{J-1}\sqrt{1+J_{\theta }J_{\theta
}^{\prime }},\quad J_{\theta }=\left\{ \left( \frac{\partial \beta }{%
\partial \theta _{1}}\right) ,...,\left( \frac{\partial \beta }{\partial
\theta _{J-1}}\right) \right\} ,
\end{equation*}%
where $\partial \beta /\partial \theta _{j}=s_{j}-s_{J}$, $j=1,2,...,J-1$.
So for any measurable set $C\subset \Theta _{\beta ,\theta }$, 
\begin{eqnarray*}
\Pr \left\{ \left( \beta ,\theta \right) \in C\right\} &=&\int_{C_{\theta
}}p(\beta ,\theta )\sqrt{1+\sum_{j=1}^{J-1}\left( \frac{\partial \beta }{%
\partial \theta _{j}}\right) ^{2}}\mathrm{d}\theta =\int_{C_{\theta
}}p(\beta ,\theta )\sqrt{1+\sum_{j=1}^{J-1}\left( s_{j}-s_{J}\right) ^{2}}%
\mathrm{d}\theta \\
&\propto &\int_{C_{\theta }}p(\beta ,\theta )\mathrm{d}\theta .
\end{eqnarray*}%
where $C_{\theta }$ is the projection of $C$ on $\theta $ (The last
proportionality is due to the fact that the Jacobian only depends on the
support of the data). Thus the linearity of the moment condition (that
results in a flat parameter space $\Theta _{\beta ,\theta }$) translates
into a somewhat trivial multiplicative correction factor and so yields a
simple relationship between $\Pr \left\{ \left( \beta ,\theta \right) \in
C\right\} $ and $p(\beta ,\theta )$.
\end{example}

\begin{example}
\label{exa:Regression} \textbf{(Regression)} The previous example can be
generalized to the family of regression models. For instance consider a
linear regression model, $\mathbb{E}\left( s^{(1)}|s^{(2)}\right) =\beta
^{\prime }s^{(2)}$, where $s=(s^{(1)},s^{(2)})$, in which $s^{(1)}$ is a
scalar and $s^{(2)}$ is a $d$-dimensional vector, and $\beta $ is a $p$%
-dimensional vector of parameters. The linear regression parameters solve
the following moment condition equation, 
\begin{equation*}
\mathbb{E}\left[ g(s,\beta )\right] =\mathbb{E}\left[ s^{(2)}(s^{(1)}-\beta
^{\prime }s^{(2)})\right] =0.
\end{equation*}%
We can also discuss the estimation of linear regression model with
instrumental variables. Assume $s=(s^{(1)},s^{(2)},s^{(3)})$, where $s^{(1)}$
is a scalar, and $s^{(2)}$ and $s^{(3)}$ are $p$-dimensional vectors
(independent and instrumental variables, respectively). If we define $%
g(s,\beta )=s^{(3)}(s^{(1)}-\beta ^{\prime }s^{(2)})$, then $\beta $ is the
solution to $\mathbb{E}\left[ g(s,\beta )\right] =0$. Moreover generalizing
to the nonlinear regression model is easy. Assume $\mathbb{E}\left(
s^{(1)}|s^{(2)}\right) =\mu (s^{(2)},\beta )$. Then the corresponding moment
condition equation is $g(s,\beta )=s^{(2)}\left\{ s^{(1)}-\mu (s^{(2)},\beta
)\right\} $. For instance for a Poisson regression $g(s,\beta
)=s^{(2)}\left\{ s^{(1)}-\exp (\beta ^{\prime }s^{(2)})\right\} $.
\end{example}

\begin{example}
\label{exa:HTEstimator} \textbf{(Average treatment effect)} Consider a
casual inference problem with the observational data $%
Z_{j}=(X_{j},Y_{j},W_{j})$ (for $1\leq j\leq N$), where $X_{j}$ is the $K$%
-dimensional vector of the $j$-th unit's background variables, $Y_{j}$ is
its scalar outcome variable, and $W_{j}$ is the binary treatment indicator.
Assuming the super-population unconfoundedness, it can be shown that (\cite%
{ImbensRubin(15)}) $\mathbb{E}_{SP}\left[ Y_{j}(1)\right] =\mathbb{E}\left[ 
\frac{W_{j}Y_{j}}{e(X_{j})}\right] $ and $\mathbb{E}_{SP}\left[ Y_{j}(0)%
\right] =\mathbb{E}\left[ \frac{(1-W_{j})Y_{j}}{1-e(X_{j})}\right] $, where $%
e(X_{j})$ is the propensity score, $e(X_{j})=\eta _{j}=\Pr (W_{j}=1|X_{j})$.
Therefore the average treatment effect (ATE) is 
\begin{equation*}
\tau =\mathbb{E}_{SP}\left[ Y_{j}(1)\right] -\mathbb{E}_{SP}\left[ Y_{j}(0)%
\right] =\mathbb{E}\left( \frac{W_{j}Y_{j}}{e(X_{j})}-\frac{(1-W_{j})Y_{j}}{%
1-e(X_{j})}\right) ,
\end{equation*}%
One might use a logistic regression model for the propensity score, $\eta
_{j}=\exp (\gamma ^{\prime }X_{j})/\left\{ 1+\exp (\gamma ^{\prime
}X_{j})\right\} $, where $\gamma $ is $K$-dimensional. Under these
assumptions the model's parameters, $\beta =(\gamma ,\tau )$, solve the
following set of moment conditions, 
\begin{equation*}
\mathbb{E}\left[ g(Z_{j},\beta )\right] =\mathbb{E}\left[ 
\begin{array}{c}
X_{j}(Y_{j}-\eta _{j}) \\ 
\frac{(W_{j}-\eta _{j})Y_{j}}{\eta _{j}\left( 1-\eta _{j}\right) }-\tau \\ 
\end{array}%
\right] =0,
\end{equation*}%
If we assume the data points are i.i.d. realizations from a discrete
distribution with finite and known support $S=\{s_{1},...,s_{j}\}$, $\Pr
(Z_{i}=s_{j})=\theta _{j}$, the moment conditions are, 
\begin{equation*}
\mathbb{E}\left[ g(Z_{j},\beta )\right] =\left[ 
\begin{array}{c}
\sum_{j=1}^{J}\theta _{j}X_{j}(Y_{j}-\eta _{j}) \\ 
\sum_{j=1}^{J}\theta _{j}\frac{(W_{j}-\eta _{j})Y_{j}}{\eta _{j}\left(
1-\eta _{j}\right) }-\tau%
\end{array}%
\right] =0.
\end{equation*}%
Thus the propensity scores and the ATE can be estimated jointly (e.g. \cite%
{McCandlessGustafsonAustin(09)}, \cite{ZiglerWattsYehWangCoullDominici(13)}
and \cite{ZiglerDominici(14)}).\newline
\end{example}

\section{Inference\label{sect:inference}\ }

\subsection{Likelihood and posterior}

Under the assumptions formulated above, the model's likelihood is 
\begin{equation*}
L(Z|\beta ,\theta )\propto \prod_{j=1}^{J}\theta _{j}^{n_{j}},
\end{equation*}%
where $n_{j}=\sum_{i=1}^{N}1(Z_{i}=s_{j})$. Note that although $\beta $ does
not appear in the likelihood explicitly, due to the constraints on $\beta $
and $\theta $, the data is informative about $\beta $.

The posterior is supported on the same set as the prior, $\Theta _{\beta
,\theta }$, and may be written as 
\begin{equation}
p(\beta ,\theta |Z)\propto p(\beta ,\theta )\left( \prod_{j=1}^{J}\theta
_{j}^{n_{j}}\right) .  \label{joint posterior}
\end{equation}%
The terms in (\ref{joint posterior}) are easy to compute for any $\left(
\beta ,\theta \right) $ in $\Theta _{\beta ,\theta }$, but the support is
defined implicitly.

\subsection{Accessing the posterior}

Inference can be carried out by sampling from the posterior distribution of
the parameters. However, in this problem, traditional simulation algorithms
will fail because the prior and the posterior of the model are supported on
a zero Lebesgue measure set (e.g. all the proposed moves of a
Metropolis-Hastings (MH) algorithm with a traditional proposal will be
rejected almost surely).

Here two solutions to this problem are given. In the first approach, called
the \textquotedblleft marginal method\textquotedblright , we will derive the
density function of the marginal of $\theta $, which has a density with
respect to the Lebesgue measure $p(\theta )$ and therefore can be processed
by conventional Monte Carlo methods. Examples include standard MCMC
algorithm and importance sampling. This is simple but comes at the cost of
having to solve for $\beta $ for each proposal. If finding $\beta $ (or
indeed all the values of $\beta $ which solve given $\theta $) is cheap then
this provides a very solid solution to the problem. \ 

In the second approach, called the \textquotedblleft joint
method\textquotedblright , we define a proposal in the space of $(\beta
,\theta )$ that assigns positive probability to $\Theta _{\beta ,\theta }$
(so, with positive probability, the proposed moves remain on the manifold $%
\Theta _{\beta ,\theta }$ and will be accepted). An MH algorithm with this
proposal is able to efficiently move in the space. This does not require us
to solve the moment conditions at all, which is extremely attractive for
difficult to solve moment condition models. \ 

\subsection{Marginal method}

Let $p(\beta ,\theta )$ be the density function of the model's prior or
posterior with respect to Hausdorff measure on $\Theta _{\beta ,\theta }$.
Proposition \ref{Prop:marginal} gives the marginal density of $\theta $ with
respect to Lebesgue measure. \ This implies that standard Monte Carlo
methods (e.g. MCMC, importance sampling, sequential importance sampling and
Hamiltonian Monte Carlo) can be used\footnote{%
We sample from the unconstrained $p(\eta )$, where $\eta _{j}=\log \left(
\theta _{j+1}/\theta _{j}\right) $, for $j=1,...,J-1$, with $\left\vert
\partial \theta /\partial \eta \right\vert =\prod_{j=1}^{J}\theta _{j}$.}. \ 

\begin{proposition}
\label{Prop:marginal}Let $p(\beta ,\theta )$ be the density function of the
prior or posterior with respect to Hausdorff measure supported on $\Theta
_{\beta ,\theta }$. Moreover, assume $p=r$ (the \textquotedblleft just
identified\textquotedblright\ case) and $\beta $ is uniquely determined by $%
\theta $, i.e. $\beta =\beta (\theta )$. Then the density function of $%
\theta $ with respect to Lebesgue measure is 
\begin{equation}
p(\theta )=\sqrt{\left\vert J_{\theta }J_{\theta }^{\prime
}+I_{p}\right\vert }\ p(\beta ,\theta ),
\end{equation}%
where 
\begin{equation}
J_{\theta }=\frac{\partial \beta }{\partial \theta ^{\prime }}=-\left\{ 
\mathbb{E}_{\theta }\left( \frac{\partial g}{\partial \beta ^{\prime }}%
\right) \right\} ^{-1}H_{\beta },\quad \text{where\quad }H_{\beta }=\left(
g_{1},...,g_{J-1}\right) -g_{J}\iota ^{\prime },
\end{equation}%
with $\iota $ being a $(J-1)$-vector of ones and 
\begin{equation*}
\mathbb{E}_{\theta }\left( \frac{\partial g}{\partial \beta ^{\prime }}%
\right) =\sum_{j=1}^{J}\theta _{j}\frac{\partial g(s_{j},\beta )}{\partial
\beta ^{\prime }}.
\end{equation*}
\end{proposition}

This proposition is a direct result of the \textquotedblleft area formula"
of \cite{Federer(69)} (see also \cite{DiaconisHolmesShahshahani(13)}) and it
can be generalized straightforwardly to the cases where for some values of $%
\theta $ there exist more than one $\beta $ by summing over the right hand
side for each solution in $\beta $.

The Jacobian\footnote{%
A Jacobian correction terms also appears in reversible jump MCMC (e.g. \cite%
{Green(95)}), when the chain is allowed to jump between models with
different number of parameters. However there the (one-to-one)
transformations are operating between spaces of the same dimension, and the
distributions in both spaces have densities w.r.t. Lebesgue measure. On the
other hand, the Jacobian in Proposition \ref{Prop:marginal} corrects for a
one-to-one mapping between spaces of different dimensions and relates two
densities that are defined w.r.t. different reference measures.} term $\sqrt{%
\left\vert J_{\theta }J_{\theta }^{\prime }+I_{p}\right\vert }$ depends on
the geometry of the parameter space $\Theta _{\beta ,\theta }$ (in other
words, it only depends on the moment conditions) and is independent of $%
p(\beta ,\theta )$. To compute this term we need to invert a $p\times p$
matrix and evaluate the determinant of a $p\times p$ matrix. However, $p$ is
usually small, in which case the computational cost of these operations is
negligible.

Importantly knowledge of the functional form of $\beta $ as a function of $%
\theta $ is not needed, since the partial derivatives can be obtained by the
implicit function theorem. However, in order to evaluate this density
function for a given $\theta $, we need its corresponding $\beta $. Although
in some problems $\beta $ has a known analytic form as a function of $\theta 
$, in many other situations it can be obtained through a numeric
optimization. We now return to the examples introduced in Section \ref%
{sect:Bayes moment}. 
\begin{figure}[tbp]
\begin{center}
\includegraphics[width=11cm]{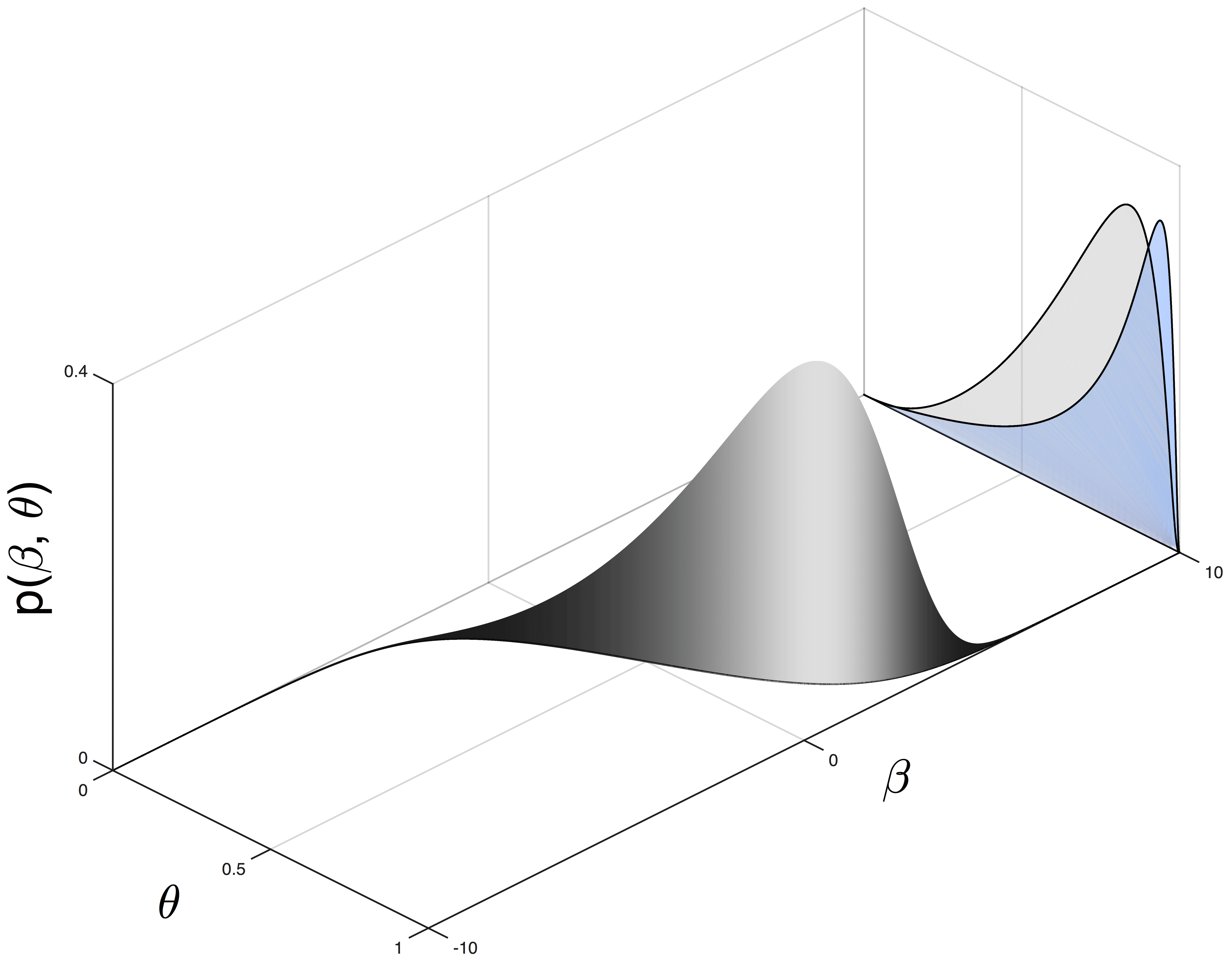}
\end{center}
\caption{Projection to the marginal density for $\protect\theta $. Blue
density is the correct marginal density $p(\protect\theta )$, given in (%
\protect\ref{p theta logit}), with respect to Lebesgue measure. The grey
density is the naive density $p\left[ \protect\beta =\log \left\{ \protect%
\theta /\left( 1-\protect\theta \right) \right\} ,\protect\theta \right] $
which ignores the corresponding length of the support. }
\label{Fig:thetabeta}
\end{figure}

\begin{example}[continues=exa:ExampleLogit1]
The density of $\theta $ in the logistic model is 
\begin{equation}
p(\theta )=p(\beta ,\theta )\sqrt{1+\left( \frac{\partial \log \left( \frac{%
\theta }{1-\theta }\right) }{\partial \theta }\right) ^{2}}.
\label{p theta logit}
\end{equation}%
Thus moment condition impacts the marginal prior on $\theta $. Figure \ref%
{Fig:thetabeta} shows the function $p(\theta )$, which has blue shade below
the curve, together with the naive $p\left( \beta =\log \left\{ \theta
/\left( 1-\theta \right) \right\} ,\theta \right) $, which has grey shade.
We can see the correct density is higher for high values of $\theta $ as
there are more dense values of $\beta $ compatible with high values of $%
\theta $ than when $\theta $ is close to 0.5.
\end{example}

\begin{example}[continues=exa:ExampleMeanJ3]
The density of $\theta $ in the mean model is 
\begin{equation*}
p(\theta )=p(\beta ,\theta )\sqrt{1+\sum_{j=1}^{J-1}\left(
s_{j}-s_{J}\right) ^{2}}\ \propto p(\beta ,\theta ).
\end{equation*}%
Hence in this case the geometry of moment condition does not impact the
prior on $\theta $. This will be the case generally when the parameter
space, $\Theta _{\beta ,\theta }$, is flat.
\end{example}

\begin{example}[continues=exa:Regression]
For the regression model write $g_{j}=g(s_{j},\beta )$ for $1\leq j\leq J$.
Therefore 
\begin{equation*}
\frac{\partial g_{j}}{\partial \beta ^{\prime }}=-s_{j}^{(2)}s_{j}^{(2)%
\prime },\quad \text{and\quad }\frac{\partial \beta }{\partial \theta _{i}}%
=\left( \sum_{j=1}^{J}\theta _{j}s_{j}^{(2)}s_{j}^{(2)\prime }\right)
^{-1}\left( g_{i}-g_{J}\right) .
\end{equation*}%
Moreover 
\begin{equation*}
J_{\theta }J_{\theta }^{\prime }=\left( \sum_{j=1}^{J}\theta
_{j}s_{j}^{(2)}s_{j}^{(2)\prime }\right) ^{-1}\left\{ \sum_{i=1}^{J}\left(
g_{i}-g_{J}\right) \left( g_{i}-g_{J}\right) ^{\prime }\right\} \left(
\sum_{j=1}^{J}\theta _{j}s_{j}^{(2)}s_{j}^{(2)\prime }\right) ^{-1}.
\end{equation*}%
Similarly for the linear regression model with instrumental variables we
have, 
\begin{equation*}
\frac{\partial g_{j}}{\partial \beta ^{\prime }}=-s_{j}^{(3)}s_{j}^{(2)%
\prime },\quad \text{and\quad }\frac{\partial \beta }{\partial \theta _{i}}%
=\left( \sum_{j=1}^{J}\theta _{j}s_{j}^{(3)}s_{j}^{(2)\prime }\right)
^{-1}\left( g_{i}-g_{J}\right) ,
\end{equation*}%
and therefore 
\begin{equation*}
J_{\theta }J_{\theta }^{\prime }=\left( \sum_{j=1}^{J}\theta
_{j}s_{j}^{(3)}s_{j}^{(2)\prime }\right) ^{-1}\left\{ \sum_{i=1}^{J}\left(
g_{i}-g_{J}\right) \left( g_{i}-g_{J}\right) ^{\prime }\right\} \left(
\sum_{j=1}^{J}\theta _{j}s_{j}^{(3)}s_{j}^{(2)\prime }\right) ^{-1}.
\end{equation*}

Again generalizing to nonlinear regression models is straightforward. If we
define $\mu _{j}=\mu (\beta ,s_{j}^{(2)})$, then 
\begin{equation*}
\frac{\partial g_{j}}{\partial \beta ^{\prime }}=-s_{j}^{(2)}\frac{\partial
\mu _{j}}{\partial \beta ^{\prime }},\quad \text{and\quad }\frac{\partial
\beta }{\partial \theta _{i}}=\left( \sum_{j=1}^{J}\theta _{j}s_{j}^{(2)}%
\frac{\partial \mu _{j}}{\partial \beta ^{\prime }}\right) ^{-1}\left(
g_{i}-g_{J}\right) ,
\end{equation*}%
which implies 
\begin{equation*}
J_{\theta }J_{\theta }^{\prime }=\left( \sum_{j=1}^{J}\theta _{j}s_{j}^{(2)}%
\frac{\partial \mu _{j}}{\partial \beta ^{\prime }}\right) ^{-1}\left\{
\sum_{i=1}^{J}\left( g_{i}-g_{J}\right) \left( g_{i}-g_{J}\right) ^{\prime
}\right\} \left( \sum_{j=1}^{J}\theta _{j}s_{j}^{(2)}\frac{\partial \mu _{j}%
}{\partial \beta ^{\prime }}\right) ^{-1}.
\end{equation*}%
For instance for $\mu (\beta ,s^{(2)})=\exp (\beta ^{\prime (2)})$ we have,%
\begin{equation*}
\frac{\partial g_{j}}{\partial \beta ^{\prime }}=-s_{j}^{(2)}\exp (\beta
_{j}^{\prime (2)})s_{j}^{(2)^{\prime }},\quad \text{and\quad }\frac{\partial
\beta }{\partial \theta _{i}}=\left( \sum_{j=1}^{J}\theta
_{j}s_{j}^{(2)}\exp (\beta _{j}^{\prime (2)})s_{j}^{(2)^{\prime }}\right)
^{-1}\left( g_{i}-g_{J}\right) ,
\end{equation*}%
and hence 
\begin{equation*}
J_{\theta }J_{\theta }^{\prime }=\left( \sum_{j=1}^{J}\theta
_{j}s_{j}^{(2)}\exp (\beta _{j}^{\prime (2)})s_{j}^{(2)^{\prime }}\right)
^{-1}\left\{ \sum_{i=1}^{J}\left( g_{i}-g_{J}\right) \left(
g_{i}-g_{J}\right) ^{\prime }\right\} \left( \sum_{j=1}^{J}\theta
_{j}s_{j}^{(2)}\exp (\beta _{j}^{\prime (2)})s_{j}^{(2)^{\prime }}\right)
^{-1}.
\end{equation*}
\end{example}

\begin{example}[continues=exa:HTEstimator]
For the casual inference problem write $g_{j}=g(s_{j},\beta )$, for $1\leq
j\leq J$. Then 
\begin{equation*}
\frac{\partial g_{j}}{\partial \beta ^{\prime }}=\left[ 
\begin{array}{cc}
s_{j}^{(1)}\eta _{j}(1-\eta _{j})s_{j}^{(1)^{\prime }} & 0_{K\times 1} \\ 
0_{1\times K} & -1 \\ 
& 
\end{array}%
\right] ,\quad \text{and\quad }\frac{\partial \beta }{\partial \theta _{i}}%
=\left( \left[ 
\begin{array}{cc}
\sum_{j=1}^{J}\theta _{j}s_{j}^{(1)}\eta _{j}(1-\eta _{j})s_{j}^{(1)^{\prime
}} & 0_{K\times 1} \\ 
0_{1\times K} & -1 \\ 
& 
\end{array}%
\right] \right) ^{-1}\left( g_{i}-g_{J}\right) ,
\end{equation*}%
which implies 
\begin{eqnarray}
J_{\theta }J_{\theta }^{\prime } &=&\left( \left[ 
\begin{array}{cc}
\sum_{j=1}^{J}\theta _{j}s_{j}^{(1)}\eta _{j}(1-\eta _{j})s_{j}^{(1)^{\prime
}} & 0_{K\times 1} \\ 
0_{1\times K} & -1 \\ 
& 
\end{array}%
\right] \right) ^{-1}\left\{ \sum_{i=1}^{J}\left( g_{i}-g_{J}\right) \left(
g_{i}-g_{J}\right) ^{\prime }\right\}  \notag \\
&&\left( \left[ 
\begin{array}{cc}
\sum_{j=1}^{J}\theta _{j}s_{j}^{(1)}\eta _{j}(1-\eta _{j})s_{j}^{(1)^{\prime
}} & 0_{K\times 1} \\ 
0_{1\times K} & -1 \\ 
& 
\end{array}%
\right] \right) ^{-1}.  \notag
\end{eqnarray}
\end{example}

An immediate consequence of Proposition \ref{Prop:marginal} is that if we
reparametrize the scientific parameters of interest $\psi =\psi (\beta )$
using a one to one transform, then 
\begin{equation}
p(\psi ,\theta )=\frac{\sqrt{\left\vert \frac{\partial \beta }{\partial
\theta ^{\prime }}\frac{\partial \beta }{\partial \theta ^{\prime }}^{\prime
}+I_{p}\right\vert }}{\sqrt{\left\vert \frac{\partial \psi }{\partial \theta
^{\prime }}\frac{\partial \psi }{\partial \theta ^{\prime }}^{\prime
}+I_{p}\right\vert }}p(\beta ,\theta ),
\end{equation}%
where $p(\psi ,\theta )$ and $p(\beta ,\theta )$ are densities with respect
to Hausdorff measures. \ 

\subsection{Joint method}

Alternatively, we may draw random samples directly from the posterior of $%
(\beta ,\theta )$. This distribution is supported on a zero Lebesgue measure
set, $\Theta _{\beta ,\theta }$, with the density function (with respect to
Hausdorff measure) $p(\beta ,\theta )$. If we ignore this and propose moves
from a continuous proposal distribution in $\mathbb{R}^{J+p-1}$ (for
instance a Gaussian proposal), the proposed moves are off the support of $%
p(\beta ,\theta )$ almost surely, and they will be rejected with probability
one. Therefore in order to sample from $p(\beta ,\theta )$ we must find a
proposal distribution that assigns positive probability to $\Theta _{\beta
,\theta }$. Drawing random samples from this proposal should be easy and
fast and (in order to compute the acceptance probability) we should be able
to evaluate its density function. This subsection will explain how this can
be achieved.

For a given value of $\beta $, the moment conditions imply the affine
constraints on $\theta $: 
\begin{equation}
H_{\beta }\theta +g_{J}=0.  \label{linear constraint}
\end{equation}%
Therefore $\Theta _{\theta |\beta }$ is a $(J-1)$-hyperplane in $\mathbb{R}%
^{J+p-1}$. This property allows us to define a suitable proposal
distribution for $(\beta ,\theta )$. Assume the current state of the MCMC is 
$(\beta ^{(t)},\theta ^{(t)})$. First we explain how a random sample from
the proposal can be drawn, and then will show how the density of this
proposal can be evaluated. In order to draw a random sample from $q(\cdot
,\cdot |\beta ^{(t)},\theta ^{(t)})$,

\begin{enumerate}
\item Draw $\beta^* | \beta^{(t)}, \theta^{(t)}$ from an (almost) arbitrary
proposal $q(\cdot | \beta^{(t)}, \theta^{(t)})$.

\item Draw $\theta ^{\ast }$ from a singular distribution supported on the
hyperplane $\mathcal{P}^{\ast }=\{\lambda \in \mathbb{R}^{J-1};H_{\beta
^{\ast }}^{\ast }\lambda +g_{J}^{\ast }=0\}$. We denote the density of this
distribution (with respect to the Hausdorff measure) by $q(\cdot |\beta
^{(t)},\theta ^{(t)},\beta ^{\ast })$. Moreover we assume the density can be
easily evaluated at any $\theta ^{\ast }$. A singular Normal distribution
supported on $\mathcal{P}^{\ast }$ is one suitable choice (see \cite%
{Khatri(68)}). In the Appendix \ref{sect:joint method proposal} we provide a
way to determine the parameters of a singular Normal distribution that can
be used to propose for $\theta ^{\ast }|\beta ^{(t)},\theta ^{(t)},\beta
^{\ast }$.
\end{enumerate}

So far we have shown how a random proposal can be generated from $q(\cdot
,\cdot |\beta ^{(t)},\theta ^{(t)})$. The following propositions
demonstrates how the density of this proposal can be evaluated when $p=r$.

\begin{proposition}
\label{prop:joint method}Let $p(\beta ,\theta )$ be the density of $(\beta
,\theta )$ with respect to $(J-1)$-dimensional Hausdorff measure on $\Theta
_{\beta ,\theta }$. Moreover assume the density of $\beta $ with respect to
Lebesgue measure is $p(\beta )$, and the density of $\theta |\beta $ with
respect to Hausdorff measure is $p(\theta |\beta )$ on $\Theta _{\theta
|\beta }$, where $\Theta _{\theta |\beta }$ is a hyperplane. Then 
\begin{equation}
p(\beta ,\theta )=\frac{\left\vert J_{\theta }J_{\theta }^{\prime
}\right\vert ^{\frac{1}{2}}}{\left\vert J_{\theta }J_{\theta }^{\prime
}+I_{p}\right\vert ^{\frac{1}{2}}}\ p(\beta )p(\theta |\beta ),\quad \text{%
where\quad }J_{\theta }=\frac{\partial \beta }{\partial \theta ^{\prime }}.
\label{joint density}
\end{equation}
\end{proposition}

The proposed pairs $(\beta ^{\ast },\theta ^{\ast })$ satisfy the moment
conditions, however the probabilities may not satisfy the probability axioms
(as some of $\theta ^{\ast }$ may be negative or $\theta _{J}^{\ast
}=1-\iota ^{\prime }\theta ^{\ast }\leq 0$). Obviously in these cases the
proposal is rejected (since the posterior is zero), the MCMC algorithm
sticks, and the proposal's density need not to be evaluated. If the proposal
is valid, then the move $(\beta ,\theta )\rightarrow (\beta ^{\ast },\theta
^{\ast })$ is accepted with probability%
\begin{equation}
\min \left\{ 1,\frac{p(\beta ^{\ast },\theta ^{\ast }|Z)}{p(\beta ,\theta |Z)%
}\frac{q(\beta ,\theta |\beta ^{\ast },\theta ^{\ast })}{q(\beta ^{\ast
},\theta ^{\ast }|\beta ,\theta )}\right\} .
\end{equation}%
The terms inside this acceptance probability are straightforward to compute
up to proportionality.

Note that in the joint method we do not need to solve for $\beta $ in each
iteration of the simulation, because our proposed moves are elements of the
parameter space $\Theta _{\beta ,\theta }$. Moreover, when $J$ goes to
infinity, the Jacobian term in (\ref{joint density}) converges to $1$. To
see this assume the data generating process is a continuous distribution or
a discrete distribution with infinite support, $s_{j}\sim H$. Then, with
probability one, just using a strong law of large numbers, 
\begin{equation}
\frac{1}{J}J_{\theta }J_{\theta }^{\prime }=\frac{1}{J}\left\{ \mathbb{E}%
_{\theta }\left( \frac{\partial g}{\partial \beta ^{\prime }}\right)
\right\} ^{-1}H_{\beta }H_{\beta }^{\prime }\left\{ \mathbb{E}_{\theta
}\left( \frac{\partial g}{\partial \beta ^{\prime }}\right) ^{\prime
}\right\} ^{-1}\rightarrow \mathcal{J}.  \notag
\end{equation}%
where $\mathcal{J=}\left\{ \mathbb{E}_{\theta }\left( \frac{\partial g}{%
\partial \beta ^{\prime }}\right) \right\} ^{-1}\mathbb{E}%
_{H}(g_{j}g_{j}^{\prime })\left\{ \mathbb{E}_{\theta }\left( \frac{\partial g%
}{\partial \beta ^{\prime }}\right) ^{\prime }\right\} ^{-1}$. Therefore 
\begin{equation*}
\frac{\left\vert J_{\theta }J_{\theta }^{\prime }\right\vert }{\left\vert
J_{\theta }J_{\theta }^{\prime }+I_{p}\right\vert }=\frac{\left\vert \frac{1%
}{J}J_{\theta }J_{\theta }^{\prime }\right\vert }{\left\vert \frac{1}{J}%
J_{\theta }J_{\theta }^{\prime }+\frac{1}{J}I_{p}\right\vert }\rightarrow 1,
\end{equation*}%
with probability one as $J$ goes to infinity. This asymptotic approximation
could be used to simplify the computation of the acceptance probability, but
otherwise does not change the substance of this section, as proposals will
be made in the same way --- directly on the manifold. \ 

\subsection{Relationship to the Bayesian bootstrap}

The \cite{Rubin(81)} \textquotedblleft Bayesian bootstrap\textquotedblright\
is at the core of \cite{ChamberlainImbens(03)}. \ We can implement our
Proposition \ref{Prop:marginal} by using their Bayesian bootstrap as a
proposal which can be reweighted to allow for informative priors on $\beta $%
. Throughout we assume $\beta $ can be solved given $\theta $.

Our generalization of \cite{ChamberlainImbens(03)} starts with the {%
Dirichlet prior }$\pi ^{\ast }(\theta )\propto \prod_{j=1}^{J}\theta
_{j}^{\alpha -1}$, $\alpha >0$. The Bayesian bootstrap then simulates from
the proposal density, 
\begin{equation}
g(\theta |Z)\propto \prod_{j=1}^{J}\theta _{j}^{n_{j}+\alpha -1}.
\end{equation}%
We assume the researcher does this $M$ times, writing the draws as $\left\{
\theta ^{(k)}\right\} _{k=1,2,...,M}$. \ For each $\theta ^{(k)}$ we assume
there is a unique $\beta ^{(k)}$ which solves the corresponding moment
conditions. \cite{ChamberlainImbens(03)} stop at this point, using this
sample as a Monte Carlo estimate of the posterior.\ 

Correcting for the geometry of the problem, the actual posterior is 
\begin{equation}
p(\theta |Z)\propto p(\beta ,\theta )\left( \prod_{j=1}^{J}\theta
_{j}^{n_{j}}\right) \left\vert J_{\theta }J_{\theta }^{\prime
}+I_{p}\right\vert ^{\frac{1}{2}}.
\end{equation}%
The resulting weights from the true posterior density with respect to the
Lebesgue measure dividing by the density from the proposal are 
\begin{equation}
w^{(k)}=\frac{p(\beta ^{(k)},\theta ^{(k)})\left\vert J_{\theta
}^{(k)}J_{\theta }^{(k)^{\prime }}+I_{p}\right\vert ^{\frac{1}{2}}}{%
\prod_{j=1}^{J}\left( \theta _{j}^{(k)}\right) ^{\alpha -1}},\quad
k=1,2,...,M,
\end{equation}%
(where $J_{\theta }^{(k)}$ is equal to $J_{\theta }$ evaluated at $(\beta
^{(k)},\theta ^{(k)})$) which normalize as $w^{(k)^{\ast }}=w^{(k)}/\left(
\sum_{k=1}^{M}w^{(k)}\right) $. An encouraging aspect of this weight is that
it does not depend on the data. \ 

In the special case where $p(\beta ,\theta )\propto \pi (\beta )\pi ^{\ast
}(\theta )$, the weights may be simply evaluated as 
\begin{equation}
w^{(k)}\propto \pi (\beta ^{(k)})\left\vert J_{\theta }^{(k)}J_{\theta
}^{(k)^{\prime }}+I_{p}\right\vert ^{\frac{1}{2}},\quad k=1,2,...,M.
\label{Gary weight}
\end{equation}

We can use these weights to estimate $\mathrm{E}\left( h(\beta )|Z\right)
\simeq \frac{1}{M}\sum_{k=1}^{M}w^{(k)^{\ast }}h(\beta ^{(k)})$. This is
importance sampling, e.g. \cite{Marshall(56)}, \cite{Geweke(89)}, \cite%
{Liu(01)}. An alternative is to resample with probability proportional to
the weight $w^{(k)}$, which delivers sampling importance resampling (SIR,
see \cite{Rubin(88)}). As with all importance samplers, the weights may
become uneven although the simplicity of the structure of the weights is
encouraging. This sampling strategy becomes appealing in the models where
the $\beta $ can be computed easily for any $\theta $, and the prior
distribution of $\beta $ is not too far from the posterior obtained from the
Bayesian bootstrap.\newline

\subsection{Missing support\label{sect:missing support}}

So far we have assumed the support of the data is known. Here we extend this
to assume the support has $J^{\ast }>J$ elements, $S^{\ast
}=(s_{1},...,s_{J^{\ast }})$, where its first $\tilde{J}=J^{\ast }-J$
elements, $\tilde{S}=(s_{1},...,s_{\tilde{J}})$, have not been observed in
the sample, while the rest of its elements $S=(s_{\tilde{J}+1},...,s_{J})$
have been observed at least once. Moreover let $\theta ^{\ast }=(\tilde{%
\theta},\theta )$, where $\tilde{\theta}$ and $\theta $ are the vector of
the probabilities of the elements of $\tilde{S}$ and $(s_{\tilde{J}%
+1},...,s_{J-1})$, and define $\theta _{J^{\ast }}=1-\sum_{j=1}^{J^{\ast
}-1}\theta _{j}^{\ast }$. We assume the missing elements of the support are
i.i.d. draws from $F_{S}$, $s_{j}\overset{iid}{\sim }F_{S}$ for $j=1,...,%
\tilde{J}$, with density $f_{S}$ with respect to Lebesgue measure. The
moment conditions are then 
\begin{equation}
\sum_{j=1}^{J^{\ast }}\theta _{j}g(s_{j},\beta )=0,
\end{equation}%
while the posterior is 
\begin{equation*}
p(\beta ,\theta ^{\ast },\tilde{S}|Z)\propto p(\beta ,\theta ^{\ast },\tilde{%
S})\left( \prod_{j=1}^{J^{\ast }}\theta _{j}^{n_{j}}\right) ,
\end{equation*}%
where $n_{j}=\sum_{i=1}^{N}1(Z_{i}=s_{j})$. Note that $n_{j}=0$ for $1\leq
j\leq \tilde{J}$, while $n_{j}$ is positive for $\tilde{J}+1\leq j\leq
J^{\ast }$.\newline
Assume the researcher expresses a prior on $(\beta ,\theta ^{\ast })|\tilde{S%
}$ with respect to the Hausdorff measure (suppressing the conditioning on $S$
for notational convenience), 
\begin{equation}
p(\beta ,\theta ^{\ast }|\tilde{S}).
\end{equation}%
Therefore 
\begin{equation}
p(\beta ,\theta ^{\ast },\tilde{S})=\left( \prod_{j=1}^{\tilde{J}}f_{S}(%
\tilde{s}_{j})\right) p(\beta ,\theta ^{\ast }|\tilde{S})
\end{equation}%
Given $\theta ^{\ast }$ and $\tilde{S}$ (and $S$), $\beta $ is uniquely
determined. Therefore the core result we need to do inference is a
generalization of Proposition \ref{Prop:marginal}: the density of the
probabilities and the missing support with respect to the Lebesgue measure
is 
\begin{equation}
p(\theta ^{\ast },\tilde{S})=\left\vert J_{\theta ^{\ast }}J_{\theta ^{\ast
}}^{\prime }+J_{\tilde{S}}J_{\tilde{S}}^{\prime }+I_{p}\right\vert ^{\frac{1%
}{2}}\left( \prod_{j=1}^{\tilde{J}}f_{S}(\tilde{s}_{j})\right) p(\beta
,\theta ^{\ast }|\tilde{S}),  \label{missing support}
\end{equation}%
where 
\begin{eqnarray}
J_{\theta ^{\ast }} &=&\frac{\partial \beta }{\partial \theta ^{\ast }}%
=\left( \sum_{j=1}^{J^{\ast }}\theta _{j}\frac{\partial g_{j}}{\partial
\beta ^{\prime }}\right) ^{-1}H_{\beta }^{\ast },\quad J_{\tilde{S}}=\frac{%
\partial \beta }{\partial \tilde{S}}=\left( \sum_{j=1}^{J^{\ast }}\theta _{j}%
\frac{\partial g_{j}}{\partial \beta ^{\prime }}\right) ^{-1}\tilde{M}, 
\notag \\
\tilde{M} &=&\left\{ \theta _{1}\left( \frac{\partial g_{1}}{\partial
s_{1}^{\prime }}\right) ,\cdots ,\theta _{\tilde{J}}\left( \frac{\partial g_{%
\tilde{J}}}{\partial s_{\tilde{J}}^{\prime }}\right) \right\} .
\end{eqnarray}%
Again this result follows from the area formula. Proposition \ref{prop:joint
method} generalizes in the same way delivering 
\begin{equation}
p(\beta ,\theta ^{\ast },\tilde{S})=\frac{\left\vert J_{\theta ^{\ast
}}J_{\theta ^{\ast }}^{\prime }+J_{\tilde{S}}J_{\tilde{S}}^{\prime
}\right\vert ^{\frac{1}{2}}}{\left\vert J_{\theta ^{\ast }}J_{\theta ^{\ast
}}^{\prime }+J_{\tilde{S}}J_{\tilde{S}}^{\prime }+I_{p}\right\vert ^{\frac{1%
}{2}}}\ p(\beta )p(\theta ^{\ast }|\beta ,\tilde{S})\left( \prod_{j=1}^{%
\tilde{J}}f_{S}(\tilde{s}_{j})\right) .
\end{equation}%
Again the Jacobian will be close to one if $J^{\ast }$ is large. \ The ratio
of $J$ to $J^{\ast }$ does not make any difference to this approximation. \ 

\begin{example}[continues=exa:ExampleMeanJ3]
Now add a single point of missing support. \ Then $J^{\ast }=4$, $\theta
^{\ast }=\left( \theta _{1},\theta _{2},\theta _{3}\right) ^{\prime }$, $J=3$
and $S^{\ast }=\{s_{1},s_{2},s_{3},s_{4}\}=\{-1,0,1,s_{4}\}$. Then $\beta
=\theta _{3}-\theta _{1}+s_{4}\theta _{4}=\theta _{3}-\theta
_{1}+s_{4}(1-\theta _{1}-\theta _{2}-\theta _{3})$. For this model 
\begin{equation}
J_{\theta ^{\ast }}=\frac{\partial \beta }{\partial \theta ^{\ast }}=\left(
-1-s_{4},-s_{4},1-s_{4}\right) \quad \text{and\quad }J_{\tilde{S}}=\frac{%
\partial \beta }{\partial \tilde{S}}=\theta _{4},
\end{equation}%
and so $J_{\theta }J_{\theta }^{\prime }=2+3s_{4}^{2}$ and $J_{\tilde{S}}J_{%
\tilde{S}}^{\prime }=\theta _{4}^{2}$. Hence, writing $\theta _{4}=1-\theta
_{1}-\theta _{2}-\theta _{3}$, 
\begin{equation}
p(s_{4},\theta ^{\ast })=\left\{ \sqrt{\left( 2+3s_{4}^{2}\right) +\theta
_{4}^{2}+1}\right\} f_{S}(s_{4})p(\beta ,\theta ^{\ast }|s_{4}).
\end{equation}
\end{example}

\section{Some potential priors\label{sect:priors}}

So far we have discussed working with any prior $p(\beta ,\theta )$ which is
defined with respect to lower dimensional Hausdorff measure supported on $%
\Theta _{\beta ,\theta }$. In this section we discuss potential ways of
selecting $p(\beta ,\theta )$. As with all prior selection there is no
uniquely good way of carrying this out. 

\subsection{A non-science prior}

From a nonparametric standpoint it is natural to build a prior from $%
p(\theta )$, e.g. Dirichlet. \ Then Proposition \ref{Prop:marginal} implies
there is a unique joint prior 
\begin{equation}
p(\beta ,\theta )=\frac{p(\theta )}{\sqrt{\left\vert J_{\theta }J_{\theta
}^{\prime }+I_{p}\right\vert }},  \label{CI prior}
\end{equation}%
which achieves this. The right hand side $p(\theta )$ is the density of $%
\theta $ with respect to Lebesgue measure, while $p(\beta ,\theta )$ is the
density of $(\beta ,\theta )$ with respect to Hausdorff measure. \ This
implies 
\begin{equation}
\Pr \left\{ \left( \beta ,\theta \right) \in C\right\} =\int_{C_{\theta
}}p(\theta )\mathrm{d}\theta .
\end{equation}%
The Dirichlet special case (\ref{CI prior}) is the implicit \cite%
{ChamberlainImbens(03)} prior on $p(\beta ,\theta )$. \ 

\subsection{A prior on the science}

Proposition \ref{prop:joint method} says that 
\begin{equation}
p(\beta ,\theta )=\frac{\left\vert J_{\theta }J_{\theta }^{\prime
}\right\vert ^{1/2}}{\left\vert J_{\theta }J_{\theta }^{\prime
}+I_{p}\right\vert ^{1/2}}p(\beta )p(\theta |\beta ).  \label{joint prior}
\end{equation}%
If we place a prior on the science $p(\beta )$ with respect to the Lebesgue
measure, then we can form a scientifically centered prior on $p(\beta
,\theta )$ by specifying a prior on $p(\theta |\beta )$ with respect to the $%
J-1-p$ dimensional Hausdorff measure. This prior sits on the hyperplane $%
\theta |\beta $ satisfying the linear constraints (\ref{linear constraint})
and the probability axioms. \ One such prior is Dirichlet subject to the
constraints. Again if $J$ gets large the Jacobian in (\ref{joint prior})
will become unimportant in practice. \ \ 

\subsection{Adhoc priors}

A more brutal approach to building a prior is to define an \textquotedblleft
initial" prior (with respect to Lebesgue measure) for $\beta $ and $\theta $
which ignores the moment condition $\eta (\beta ,\theta )$ where the implied
initial marginal prior on $\beta $, $\eta \left( \beta \right) $, could be
our substantive initial prior. From the Borel paradox (\cite{Kolmogorov(56)}%
) we know there are many ways of building a $p(\beta ,\theta )$ from $\eta
(\beta ,\theta )$ (conditioning on satisfying the moment condition is not
enough) but here we discuss various plausible methods.

This line of thinking leads to a generalization of (\ref{CI prior}), setting%
\begin{equation}
p(\beta ,\theta )\propto \frac{\eta (\beta ,\theta )}{\left\vert J_{\theta
}J_{\theta }^{\prime }+I_{p}\right\vert ^{1/2}}1_{\Theta _{\beta ,\theta
}}(\beta ,\theta ).  \label{Gary prior}
\end{equation}%
This prior scales the initial prior to countereffect the length of the curve
mapping out the relationship between $\theta $ and $\beta $ implied by the
moment condition. \ This prior has the property that $p(\theta )\propto \eta
(\beta ,\theta )1_{\Theta _{\beta ,\theta }}(\beta ,\theta )$, with respect
to the Lebesgue measure.

The simple case of $\eta (\beta ,\theta )=\eta (\beta )\eta (\theta )$,
would imply under (\ref{Gary prior}) \ 
\begin{equation}
p(\theta )\propto \eta (\beta )\eta (\theta )1_{\Theta _{\beta ,\theta
}}(\beta ,\theta ).
\end{equation}%
The case where $\eta (\theta )$ is Dirichlet is important. Then the Bayesian
bootstrap weights (\ref{Gary prior}) would become the rather simple 
\begin{equation}
w_{j}\propto \eta (\beta ^{(j)}),\quad j=1,2,...,M.
\end{equation}%
This is a minimally informative generalization of \cite%
{ChamberlainImbens(03)}. \ 
\begin{figure}[tbp]
\begin{center}
\includegraphics[width=11cm]{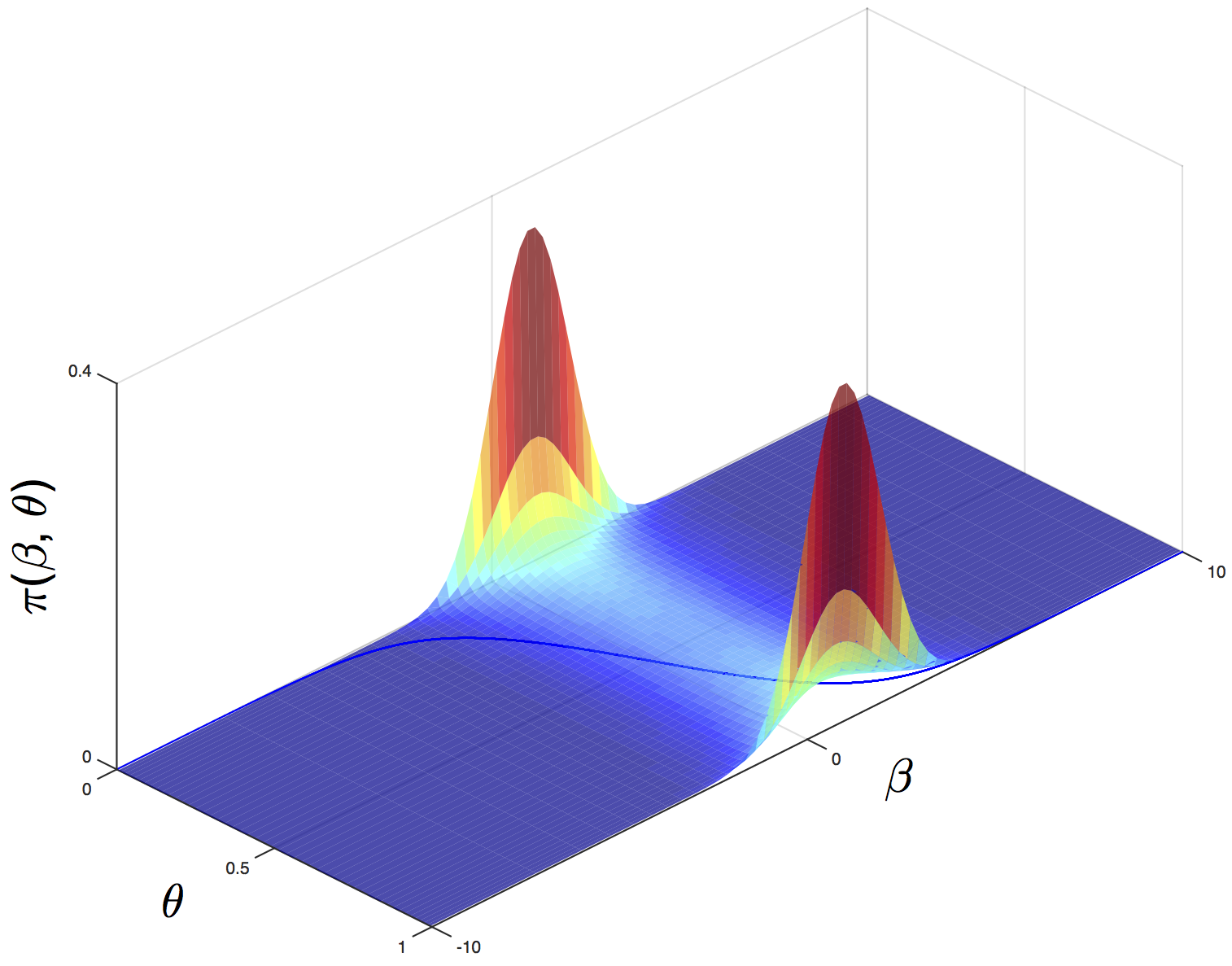}
\end{center}
\caption{The parameter space (the blue curve) $\Theta _{\protect\beta ,%
\protect\theta }$ and the initial prior $\protect\pi (\protect\beta ,\protect%
\theta )$ have been depicted. Figure \protect\ref{Fig:support_density} shows
the implied $p(\protect\beta ,\protect\theta )$.}
\label{fig:Example_Logit_InitialPrior_Prior}
\end{figure}

An alternative to (\ref{Gary prior}) is to put no mass on inadmissible
combinations of $\beta ,\theta $. We call this the \textquotedblleft
truncated prior\textquotedblright\ 
\begin{equation}
p(\beta ,\theta )\propto \eta (\beta ,\theta )1_{\Theta _{\beta ,\theta
}}(\beta ,\theta )  \label{truncated prior}
\end{equation}%
in which $p(\beta ,\theta )$ is the density of the prior with respect to the 
$(J-1)-$dimension Hausdorff measure in $\mathbb{R}^{J-1+p}$. This would
imply for any set $C\in \mathbb{R}^{J-1+p}$%
\begin{eqnarray}
\Pr \left\{ \left( \beta ,\theta \right) \in C\right\} &=&\int_{C_{\theta
}}p(\beta ,\theta )\sqrt{\left\vert J_{\theta }J_{\theta }^{\prime
}+I_{p}\right\vert }\mathrm{d}\theta \\
&\propto &\int_{C_{\theta }}\eta (\beta ,\theta )\sqrt{\left\vert J_{\theta
}J_{\theta }^{\prime }+I_{p}\right\vert }\mathrm{d}\theta .  \notag
\end{eqnarray}%
Obviously it implies $p(\theta )\propto \eta (\beta ,\theta )\sqrt{%
\left\vert J_{\theta }J_{\theta }^{\prime }+I_{p}\right\vert }1_{\Theta
_{\beta ,\theta }}(\beta ,\theta )$, with respect to the Lebesgue measure. \ 

\begin{example}
\label{exa:Example_Logit copy(1)} (continuing logistic Example \ref%
{exa:ExampleLogit1}). \ Assume the initial prior 
\begin{equation}
\eta (\beta ,\theta )\propto \theta ^{0.01-1}(1-\theta )^{0.01-1}e^{-\frac{1%
}{2}\left( \beta -1\right) ^{2}},
\end{equation}%
which is a relatively ignorant Dirichlet prior on the probabilities and an
informative Gaussian prior for $\beta $ centered on one. This is depicted in
Figure \ref{fig:Example_Logit_InitialPrior_Prior}. With this initial prior
and using the class of priors (\ref{truncated prior}), the density with
respect to the univariate Hausdorff measure is 
\begin{equation}
p(\beta ,\theta )\propto \eta (\beta ,\theta )1_{\Theta _{\beta ,\theta
}}(\beta ,\theta ).
\end{equation}%
Figure \ref{Fig:support_density} shows the corresponding $p(\beta ,\theta )$
living on the manifold. In this case%
\begin{equation}
p(\theta )\propto \eta (\beta ,\theta )\sqrt{1+\left( \frac{\partial \beta }{%
\partial \theta }\right) ^{2}}1_{\Theta _{\beta ,\theta }}(\beta ,\theta ),
\end{equation}%
with respect to the Lebesgue measure. With the alternative (\ref{Gary prior}%
) prior, then 
\begin{equation}
p(\beta ,\theta )\propto \frac{\eta (\beta ,\theta )}{\sqrt{1+\left( \frac{%
\partial \beta }{\partial \theta }\right) ^{2}}}1_{\Theta _{\beta ,\theta
}}(\beta ,\theta ),\quad \text{and\quad }p(\theta )\propto \eta (\beta
,\theta )1_{\Theta _{\beta ,\theta }}(\beta ,\theta ).
\end{equation}
\end{example}

\section{Illustrative examples\label{sectio:examples}}

In this section we present some illustrative examples and simulation
studies. Since the MCMC results obtained by the marginal and joint methods
are indistinguishable we present only one of them. At the end of the section
we study how the methods scale. \ 

\subsection{The mean}

Recall inference on the mean studied in Example \ref{exa:ExampleMeanJ3}. Now
focus on $J=3$ and $S=(-1,0,1)$, so $\beta =\theta _{3}-\theta
_{1}=1-2\theta _{1}-\theta _{2}$. Here we have taken the 2 dimensional
Hausdorff prior as 
\begin{equation}
p(\beta ,\theta )\propto e^{-2\left\vert \beta -m\right\vert }\theta
_{1}^{\alpha -1}\theta _{2}^{\alpha -1}\left( 1-\theta _{1}-\theta
_{2}\right) ^{\alpha -1}1\left( {\min \{\theta _{1},\theta _{2},1-\theta
_{1}-\theta _{2}\}\geq 0}\right) .
\end{equation}%
We call this a \textquotedblleft Laplace-Dirichlet\textquotedblright\
distribution on $\left( \beta ,\theta \right) $, where $\beta $ is centered
around $m$ and the Dirichlet part is indexed by $\alpha $.

\begin{figure}[tbp]
\includegraphics[width=16cm]{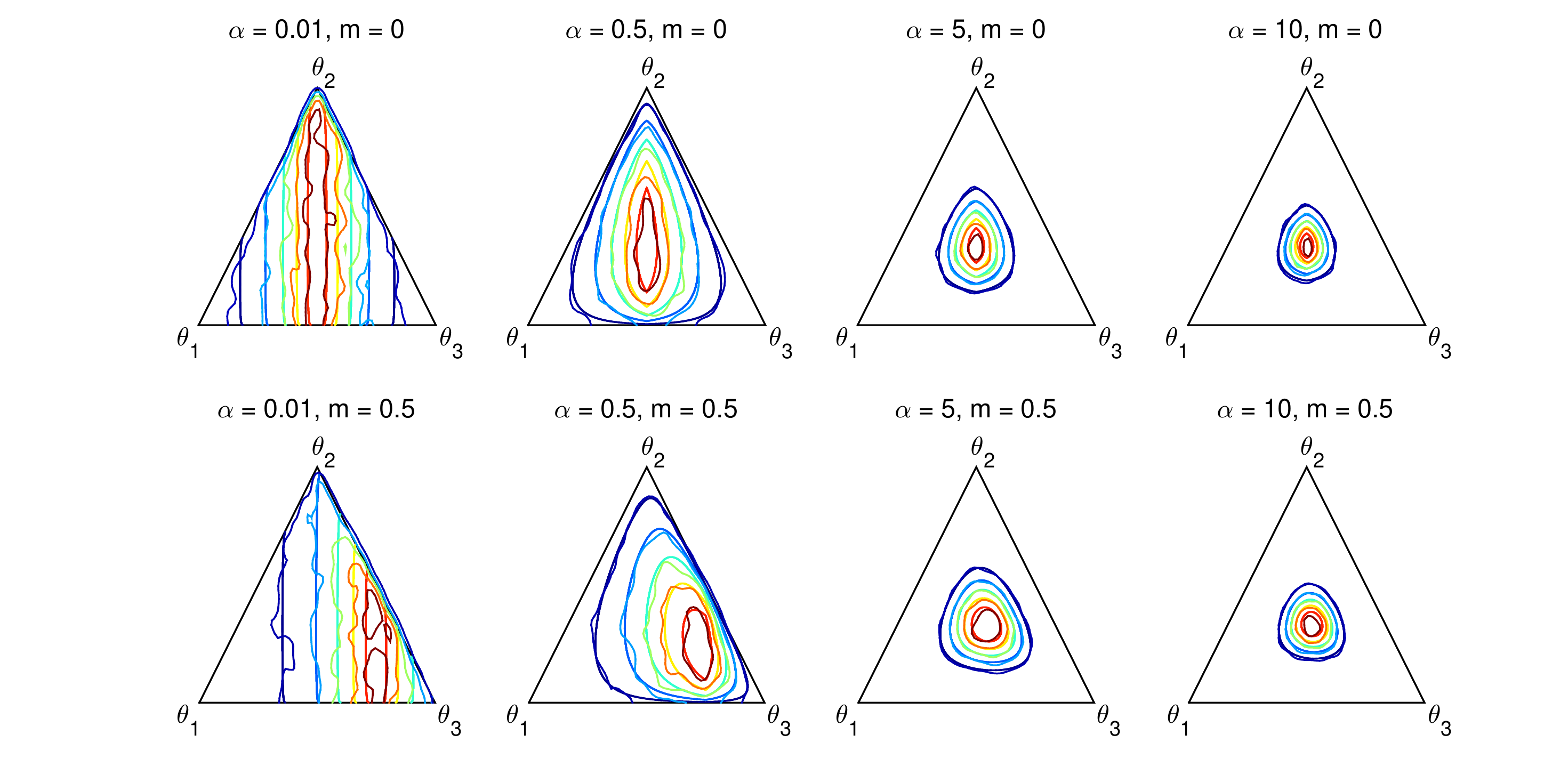} {\ }
\caption{Equiprobability contours implied by the Laplace-Dirichlet prior on $%
p(\protect\beta ,\protect\theta )$\ with respect to the Hausdorff measure. \
Plotted is the marginal $p(\protect\theta _{1},\protect\theta _{3})$\ for
several values of $m$\ and $\protect\alpha $, with $\protect\theta _{2}$\
implied as $\protect\theta _{2}=1-\protect\theta _{1}-\protect\theta _{3}$.
\ This case has $J=3$\ points of support ($s_{1}=-1,s_{2}=0,s_{3}=1$) and $%
r=1$\ moment constraints (the mean). \ \textit{\ }}
\label{fig:Example1_theta}
\end{figure}

By the marginal method: 
\begin{equation}
p(\theta )\propto e^{-2\left\vert 1-2\theta _{1}-\theta _{2}-m\right\vert
}\theta _{1}^{\alpha -1}\left( 1-\theta _{1}-\theta _{2}\right) ^{\alpha
-1}\theta _{2}^{\alpha -1}1\left( {\min \{\theta _{1},\theta _{2},1-\theta
_{1}-\theta _{2}\}\geq 0}\right) ,
\end{equation}%
Figure \ref{fig:Example1_theta} shows the contours of $p(\theta )$ for
various values of $m$ and $\alpha $. We have plotted these contours against $%
\left( \theta _{1},\theta _{2},\theta _{3}\right) ^{\prime }$ so the reader
can compare $\theta _{1}$ and $\theta _{3}$. 

If the Laplace-Dirichlet distribution has $m=0$ then the density is
symmetric with respect to $\theta _{1}$ and $\theta _{3}$. When the location
parameter of $p(\beta )$ is positive $\theta _{1}$ is on average smaller
than $\theta _{3}$. Moreover as $\alpha $ increases, the variability of $%
p(\theta )$ decreases.

\begin{figure}[tbp]
\includegraphics[width=16cm]{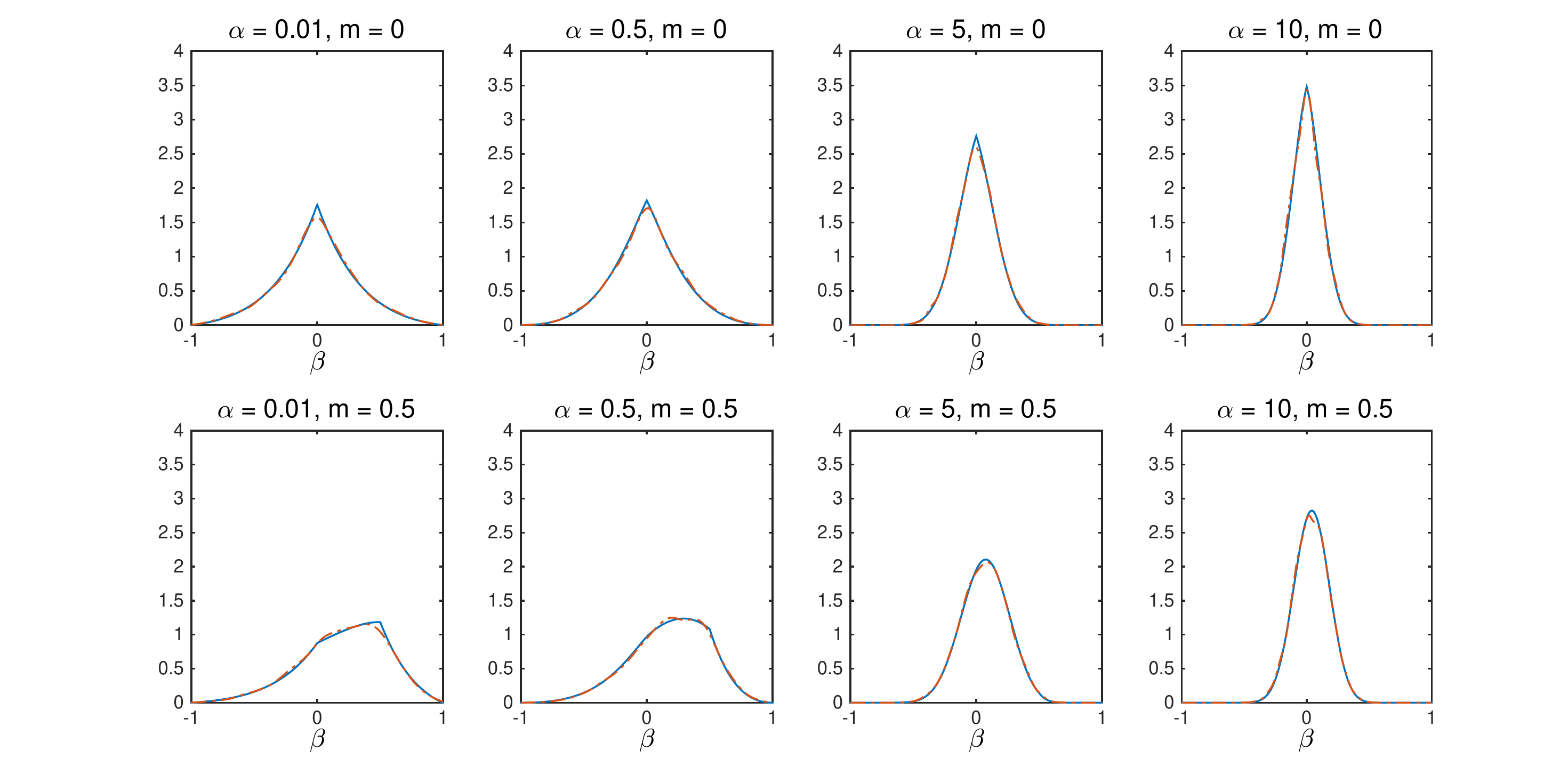}
\caption{Illustrating Example 1 (estimating the mean). Plot of $p(\protect%
\beta )$ for several values of $m$ and $\protect\alpha $. This case has $J=3$%
\ points of support ($s_{1}=-1,s_{2}=0,s_{3}=1$) and $r=1$ moment
constraints (the mean). Initial prior for $\protect\beta $ is Laplace
centered at $m$ and the initial prior for $\protect\theta $ is symmetric
Dirichlet with parameter $\protect\alpha $.}
\label{fig:Example1_beta}
\end{figure}

Figure \ref{fig:Example1_beta} draws the prior for $\beta $. Here the
support\ of the data means $\beta $ is restricted to the real line, after
observing the support of the data its prior is restricted to $[-1,1]$. As $%
\alpha $ increases the variance of $\beta $ decreases. For instance the $%
\beta $'s prior centered at a positive value results in a prior for $\theta $
tilted toward $\theta _{3}$, even if the prior of $\theta $ is symmetric. In
the same way, a more informative initial prior for $\theta $ yields a more
peaked prior for $\beta $. 

\subsection{Missing support and the mean}

In the previous section, the finite support of $\beta $ is caused by the
known support of the data. \ We now extend this to cover Example \ref%
{exa:ExampleMeanJ3} where we have a single missing datapoint 
\begin{equation}
s_{4}\sim N(0,10^{2}),  \label{marginal missing 1}
\end{equation}%
all other features of the problem are unchanged. An adaptive MH algorithm
has been used in order to draw $10,000$ samples from the joint distribution.
For the sake of brevity we present the results only for the case of $m=0$
and $\alpha =0.5$.

\begin{figure}[tbp]
\begin{center}
\includegraphics[width=12cm]{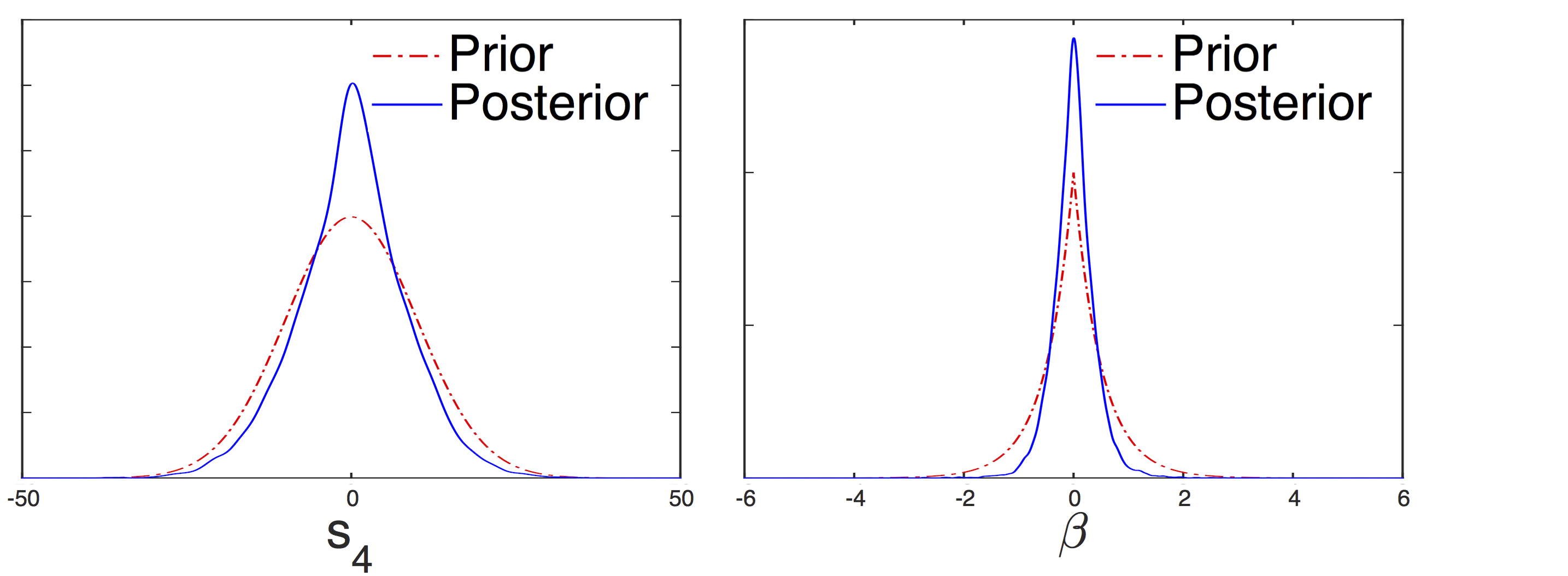}
\end{center}
\caption{Illustrating Example 1 with the partially observed support:
inferring the mean $\protect\beta $. The prior and the posterior of the
missing element of the support $s_{4}$ (left panel) and mean $\protect\beta $
(right panel).}
\label{fig:Example1_RandomSupport}
\end{figure}

The means and standard error of the probabilities $\theta ^{\ast }$ are $%
(0.3065,0.3335,0.3079,0.0521)$ and $(0.0026,0.0030,0.0024,0.0010)$,
respectively. The left panel of Figure \ref{fig:Example1_RandomSupport}
shows the initial prior (\ref{marginal missing 1}) and the implied marginal
distribution of the missing element of the support $s_{4}$ from the joint
prior. The variance of the implied marginal is smaller than the prior's
variance, because the prior distribution of $\beta $ is informative about
the support of the data. The right panel of Figure \ref%
{fig:Example1_RandomSupport} shows the Laplace element of the prior $%
e^{-2\left\vert \beta -m\right\vert }$ and the full marginal prior for $%
\beta $. \ The full marginal prior is not the same as the Laplace
distribution due to the informative priors on the probabilities. \ 

\subsection{Linear regression}

\begin{figure}[tbp]
\begin{center}
\includegraphics[width=8cm]{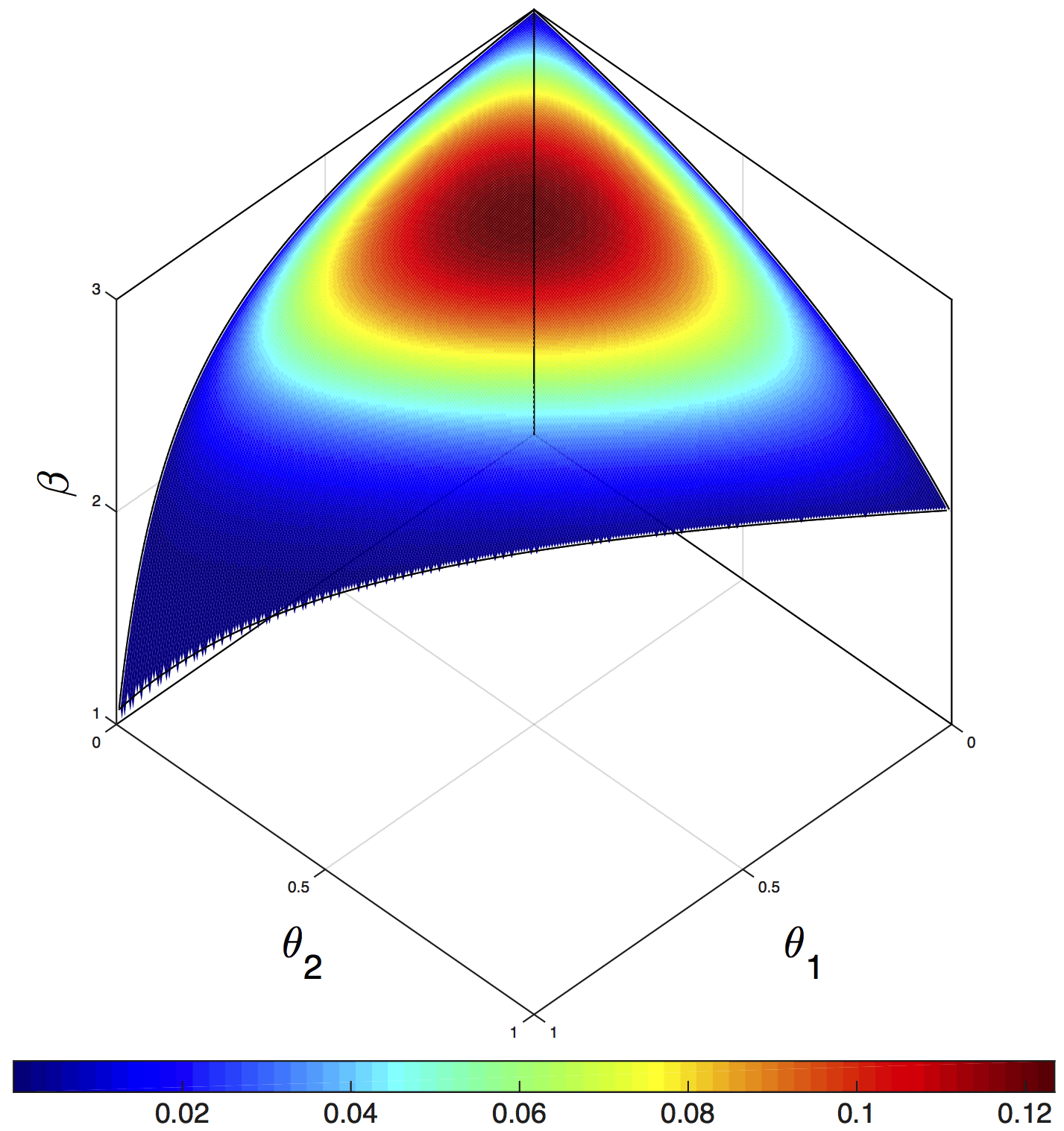}
\end{center}
\caption{The posterior distribution of the linear regression model with data 
$Z=\{(1,1),(2,4),(3,9)\}$. The prior parameters are $\protect\alpha =0.5$
and $m=3$.}
\label{fig:Example_LRWoCJ3_Posterior}
\end{figure}

Recall the linear regression of Example 3. Assume the observed data is $%
Z=\{(1,1),(2,4),(3,9)\}$. Earlier we have seen that the parameter space, $%
\Theta _{\beta ,\theta }$ is a non-flat surface in $\mathbb{R}^{3}$. Figure %
\ref{fig:Example_LRWoCJ3_Posterior} demonstrates the posterior distribution
of the parameters defined on this surface (the prior parameters are $\alpha
=0.5$ and $m=3$). 

Following the suggested MCMC simulation algorithms we draw $100,000$ samples
from the posterior distribution of the parameters. In the Figure \ref%
{fig:Example_LRWoCJ3_Contour} we have drawn the contour plots of the
posterior distribution of the probabilities. Analytical results have been
compared with the estimates obtained by a kernel density estimator using the
MCMC draws. 
\begin{figure}[tbp]
\begin{center}
\includegraphics[width=16cm]{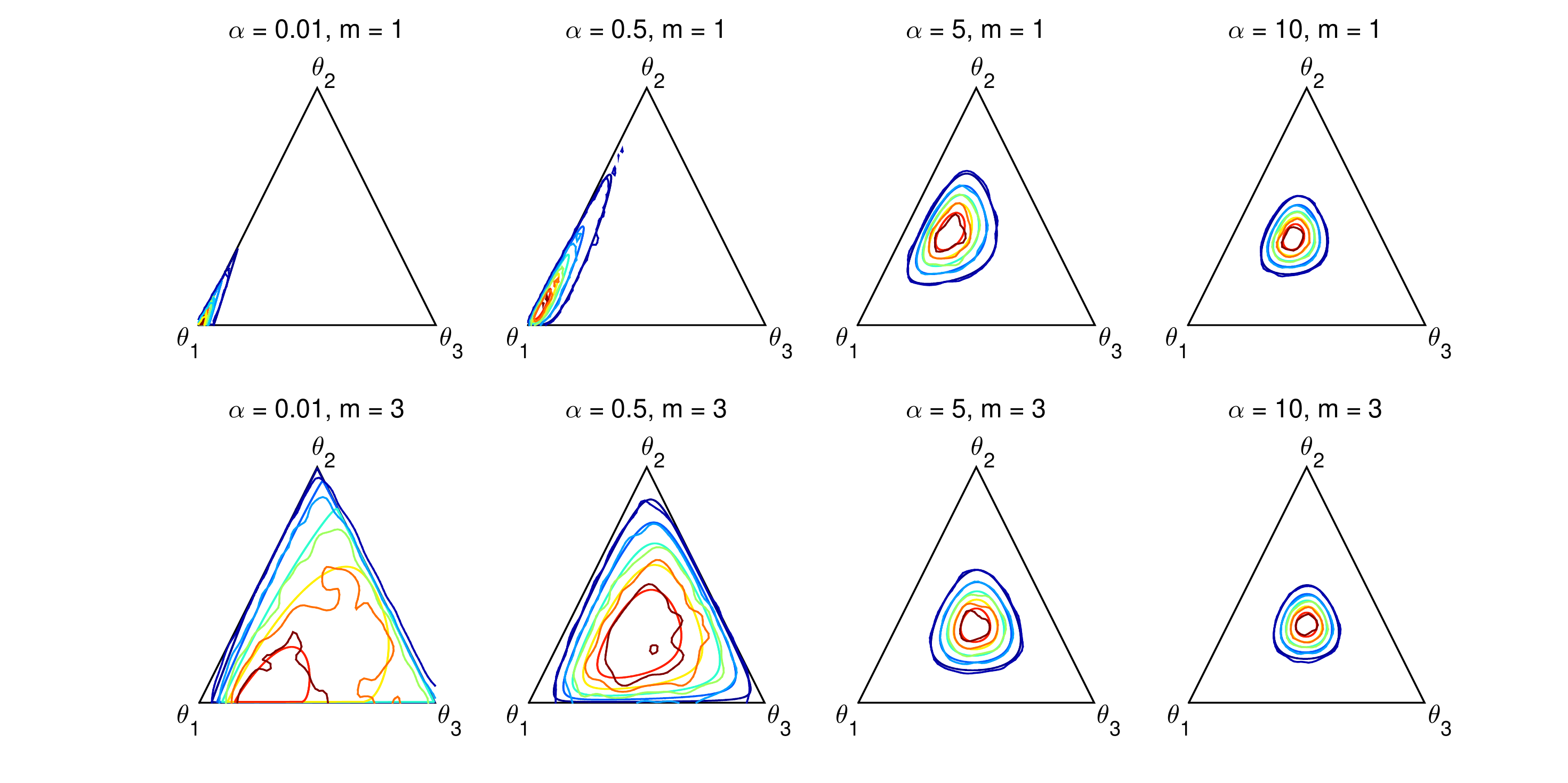}
\end{center}
\caption{The posterior distribution of $\protect\theta $ in the linear
regression model with data $Z=\{(1,1),(2,4),(3,9)\}$ (analytical results and
the estimates obtained by a kernel density estimator using $100,000$ MCMC
draws). The prior parameters are $\protect\alpha =0.5$ and $m=3$.}
\label{fig:Example_LRWoCJ3_Contour}
\end{figure}

\subsection{Simulation study\label{sect:simulation studies}}

To demonstrate the scalability of the algorithms we consider a linear
regression model with sample size $J=500$. The data $Z_{j}=(Y_{j},X_{j})$,
for $1\leq j\leq J$, is generated according to $X_{j}\sim \mathcal{N}%
(1,2^{2})$, $Y_{j}|X_{j}\sim \mathcal{N}(2+5X_{j},10^{2})$. \ We assume the
substantive prior of $\beta $ is $\beta \sim \mathcal{N}(\mu _{0},\Sigma
_{0})$, where the elements of $\mu _{0}$ are equal to the $25\%$ quantiles
of the asymptotic MLE estimators, and $\Sigma _{0}$ is equal to the
asymptotic variance of the MLE estimator multiplied by $100$ (see Appendix %
\ref{sect:appendix linear regression} for the results with a different
prior). The initial prior of $\theta $ is a symmetric Dirichlet distribution
with parameter $\alpha =0.01$. We have drawn $50,000$ samples from the
posterior after a $5,000$ sample burn-in (the chain's trace has been thinned
with a factor of $100$, so has been iterated $5,000,000$ times). The scatter
plot of the sample is depicted in the top-left panel of Figure \ref%
{fig:Example_LRJ500_Plots}. Each circle represents a data point in our
sample and its radius is proportional to the expected value of its posterior
probability, i.e. $\mathbb{E}(\theta _{j}|Z)$. In the top-right panel the
correlogram (ACF) of the chains of $\beta $ and $10$ elements of $\theta $
have been presented (the red dashed lines and the blue dotted lines are
corresponding to $\beta $ and $\theta $, respectively.) The ACFs demonstrate
that the Markov chain is mixing sufficiently well. In the bottom-left panel
the contour plot of the posterior distribution of $\beta $ has been compared
to the one obtained by the Bayesian bootstrapping of \cite%
{ChamberlainImbens(03)}. The posterior distributions are very close, because
the prior's information is roughly $1\%$ of the information content of the
sample. The bottom-right panel shows a histogram of the samples from the
posterior distribution of $\beta $.

%
%

\begin{figure}[tbp]
\begin{center}
\includegraphics[width=17cm]{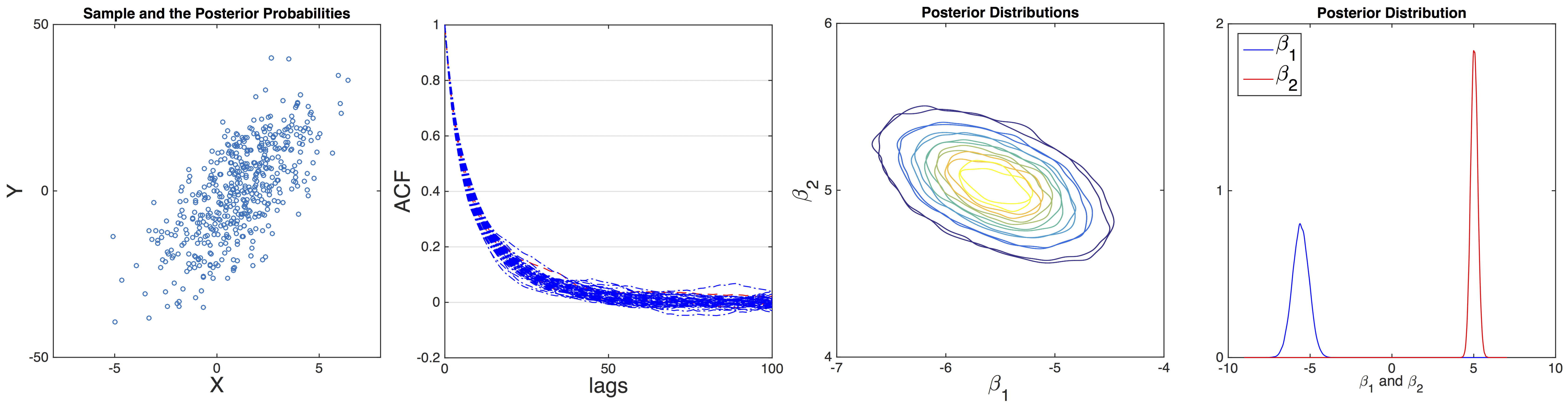}
\end{center}
\caption{Inference in linear regression model with $J=500$. Top left shows
circles who's radius is the posterior expectation of the probabilities given
the data: $E(\protect\theta _{j}|Z)$. Top right is the correlogram for the
thinned draws of the elements of $\protect\beta $ and ten elements of $%
\protect\theta $. The bottom graphs show the estimated contour and marginal
densities of the resulting posterior. }
\label{fig:Example_LRJ500_Plots}
\end{figure}

\section{Empirical studies \label{sectio: empirical studies}}

In this section we study two empirical examples. The first focuses on an
instrumental variable based estimator, the second looks at estimating the
average treatment effect from an experiment. \ 

\subsection{Instrumental variables \label{sect:empirical studies_IV}}

In this section we demonstrate the applicability and scalability of the
methodology developed in this paper to a real dataset. We use a subsample of
the earnings and schooling dataset studied in \cite{ChamberlainImbens(03)}.
This dataset is a subset of the data studied in \cite{AngristKrueger(91)}
and consist of the self-reported weekly log-earnings (self-reported annual
earnings divided by 52) of $162,512 $ male subjects who reported positive
annual wages in 1979 along with their number of years of education and their
quarter of birth date. In turn this is a 5\% random sample from the 1980
Public Use Census Data. \ \cite{BoundBrownMathiowetz(01)} discuss the myriad
of problems of self-report income data but we do not address that issue
here. \ For example, \cite{BrittonShephardVignoles(15)} compared UK
self-reported income with tax based administrative data finding high income
earners significantly under self-recorded their incomes compared to that
seen in administrative data. \ 
\begin{figure}[tbp]
\begin{minipage}[t]{0.4\linewidth}
\centering
\includegraphics[width=8cm]{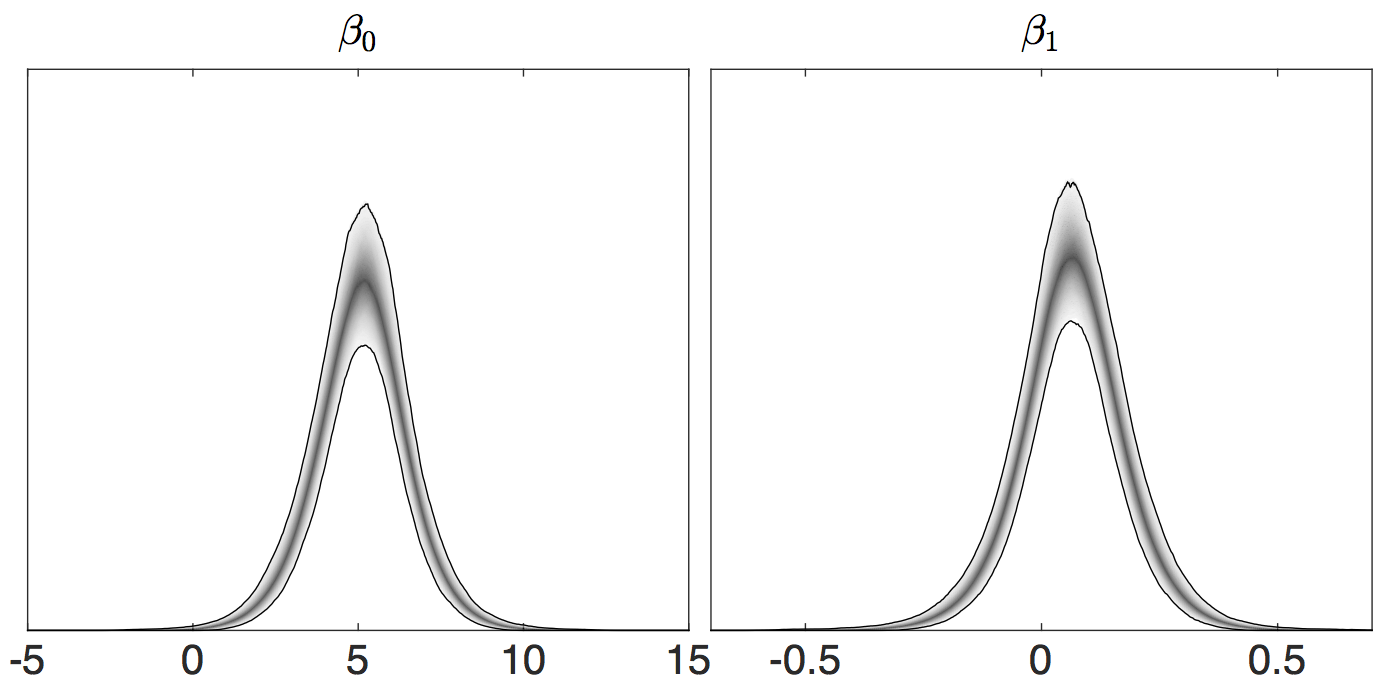} %
{\footnotesize $J=100$}
\end{minipage}
\hspace{2.0cm} 
\begin{minipage}[t]{0.4\linewidth}
\centering
\includegraphics[width=8cm]{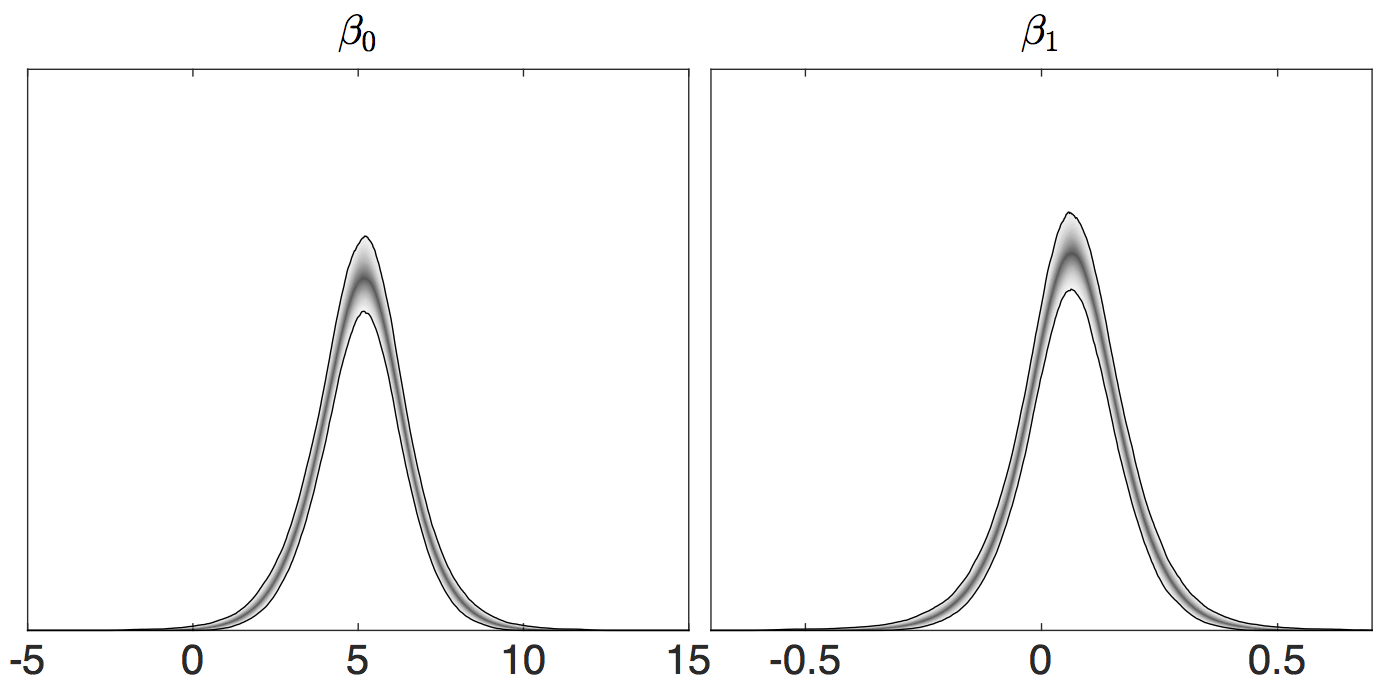} %
{\footnotesize $J=1,000$ }
\end{minipage} \\[10pt]
\begin{minipage}[t]{0.4\linewidth}
\centering
\includegraphics[width=8cm]{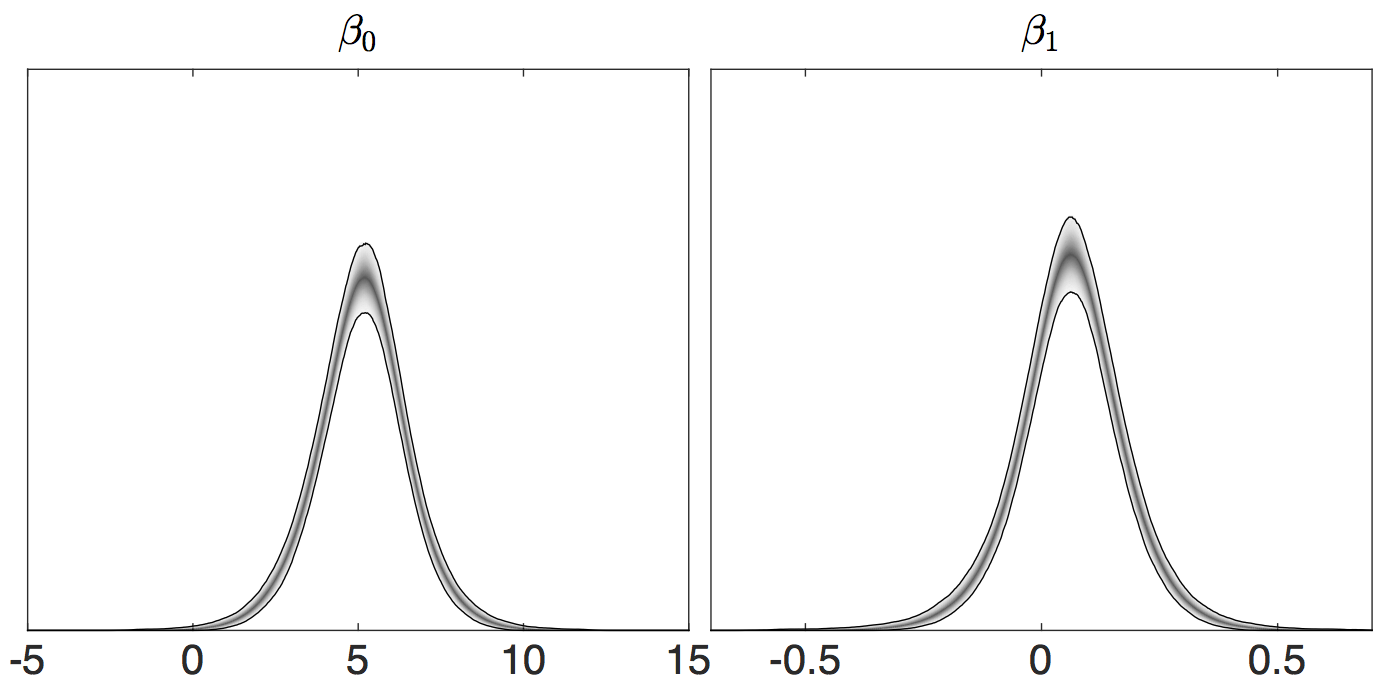} %
{\footnotesize $J=5,000$}
\end{minipage}
\hspace{2.0cm} 
\begin{minipage}[t]{0.4\linewidth}
\centering
\includegraphics[width=8cm]{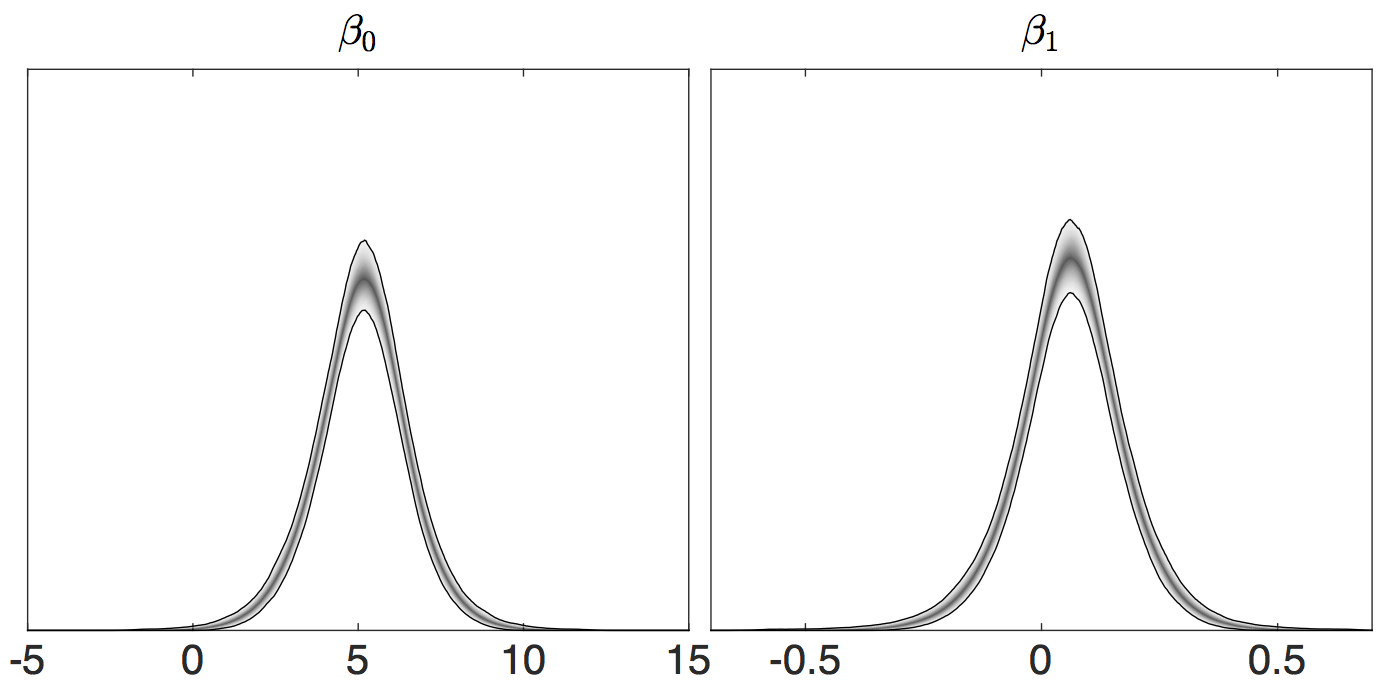} %
{\footnotesize $J=10,000$}
\end{minipage}
\caption{95\% pointwise confidence regions for the marginal prior for $%
\protect\beta $ for $J=10$, $J=1,000$, $J=10,000$ and $J=100,000$ points of
random support in top-left, top-right, bottom-left and bottom-right,
respectively. The confidence region is computed over 1,000 replications. }
\label{fig:Example_IV_MarginalPrior1}
\end{figure}

\cite{ChamberlainImbens(03)} studied the dependence of earnings on the level
of schooling using a linear additive treatment effect model (e.g. \cite%
{ImbensRubin(15)}). They model schooling levels as being determined by
rational agents' optimization of their lifetime expected utility. Since the
utility is a function of the earnings they needed to estimate the
distribution of earnings as a function of the schooling level.

The expected log-earnings $Y_{X}$ with schooling level $X$ is modeled here
as $\mathrm{E}(Y_{X}|X,Y_{0})=Y_{0}+\beta _{1}X$, where $X$ is the schooling
level, $\beta _{1}$ is the unknown return to education, and $Y_{0}$ is the
earnings level with no schooling at all. Let $\beta _{0}$ be the expected
value of $Y_{0}$, so $Y_{0}-\beta _{0}$ has a zero mean.

In order to estimate the unknown parameters, $\beta =(\beta _{0},\beta _{1})$%
, we follow \cite{AngristKrueger(91)} and \cite{ChamberlainImbens(03)} and
use an instrumental variable (IV) $W$ that is a binary indicator: $W=0$ if
the subject was born in the first three quarters of the year and $W=1$
otherwise. The instrumental variable $W$ is correlated with the regressor $X$
and thought by the researchers to be uncorrelated with the errors.

We obtain the classic IV estimate of $\beta $ using the full sample, and
treat them as the \textquotedblleft true" values of $\beta $. Then we draw
random samples with replacement of size $J$ from the original data $1,000$
times. Our aim will be to compare different estimators using these smaller
samples.\ 

Our prior distribution, which is specified to be weakly informative, is 
\begin{equation}
p(\beta ,\theta )\propto \frac{1}{\sqrt{J_{\theta }J_{\theta }^{\prime
}+I_{2}}}\eta (\beta )\eta (\theta )1_{\Theta _{\beta ,\theta }}(\beta
,\theta ),
\end{equation}%
where $\eta (\beta )=\varphi \left( \beta _{0};5,4\right) \varphi \left(
\beta _{1};0,0.2\right) $, and $\varphi \left( \cdot ;\mu ,\sigma
^{2}\right) $ is the Gaussian density with mean $\mu $ and variance $\sigma
^{2}$. The intercept is centered at $5$ with variance $4$, implying that the
mean annual income for those with no schooling is equal to $\$7,717$ (with $%
95\%$ confidence interval $[\$153,\$388965]$) with zero years of schooling.
Moreover the prior of $\beta _{1}$ has zero mean (no effect of number of
schooling years on income) with $95\%$ interval $[-0.88,0.88]$ (that is
equivalent to $[-0.41\%,241\%]$ income increment for each additional year of
schooling.) The probabilities $\theta $ are taken as a mildly informative
Dirichlet prior $\eta (\theta )\propto \dprod\limits_{j=1}^{J}\theta
_{j}^{\alpha -1}$, where $\alpha =10^{-6}$ (we also tried $\alpha =J^{-1}$,
with no substantial change in the results). 
\begin{figure}[tbp]
\begin{minipage}[t]{0.4\linewidth}
\centering
\includegraphics[width=8cm]{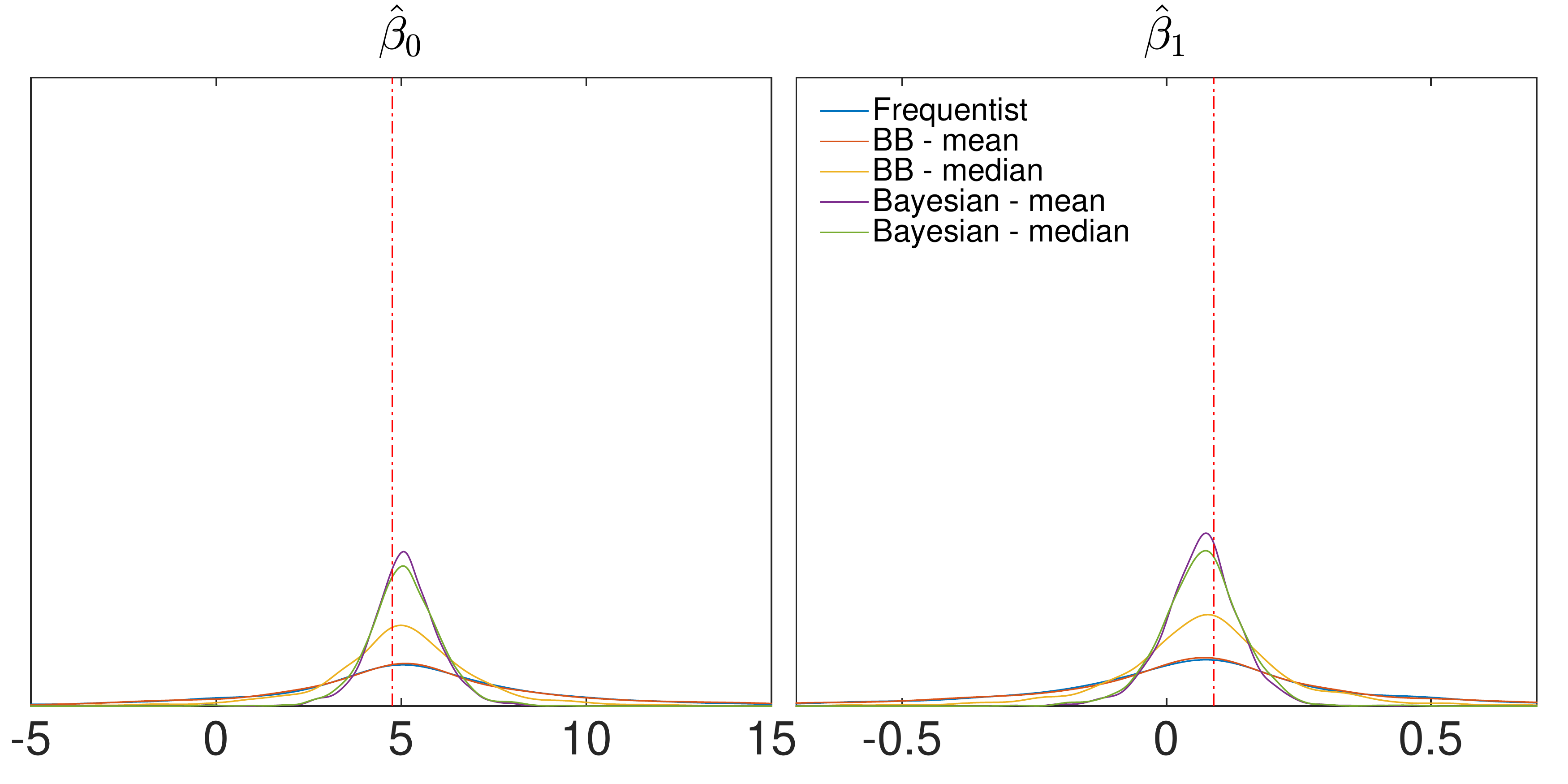} %
{\footnotesize $J=10$ }
\end{minipage}
\hspace{2.0cm} 
\begin{minipage}[t]{0.4\linewidth}
\centering
\includegraphics[width=8cm]{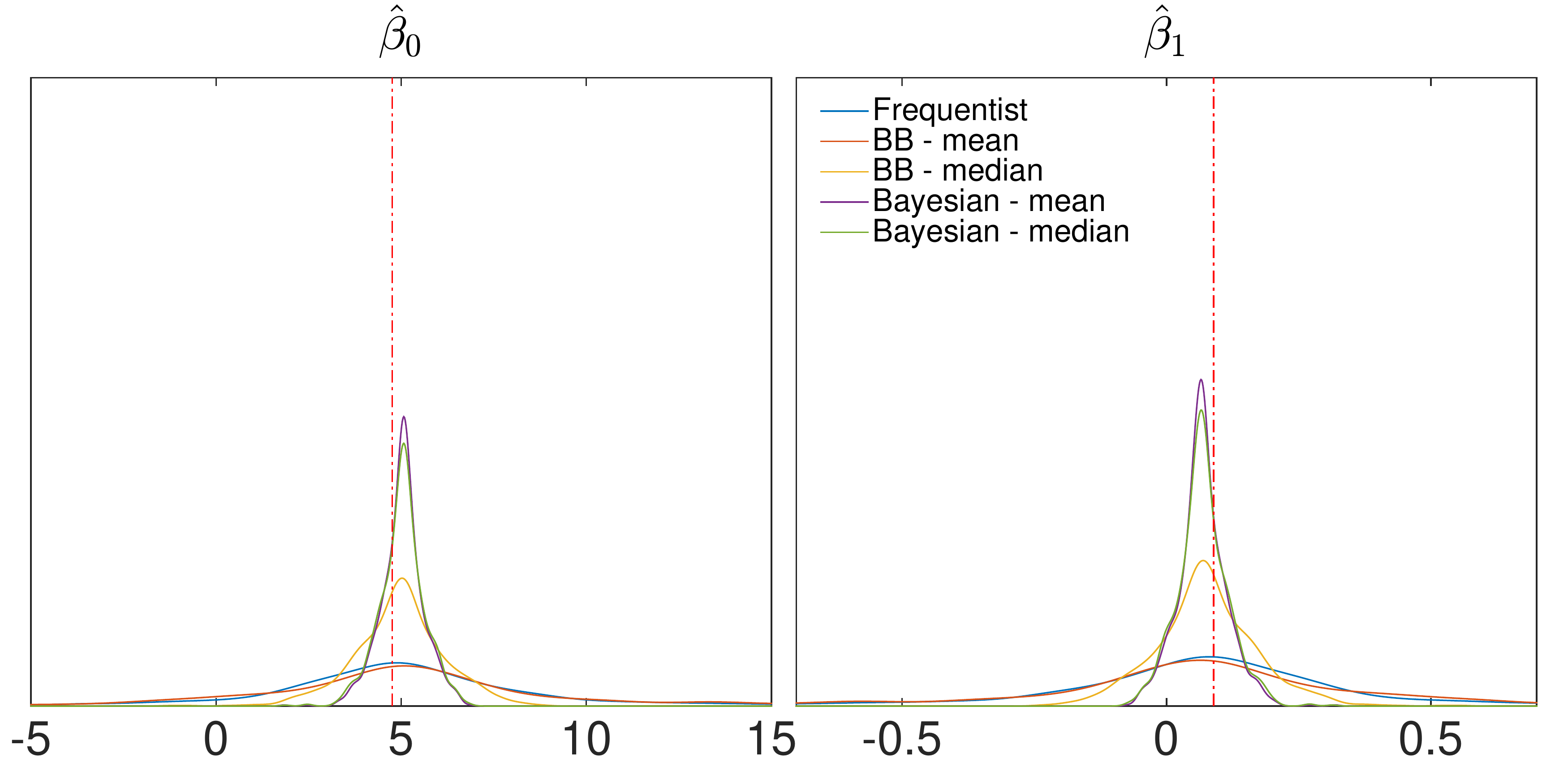} %
{\footnotesize $J=1,000$}
\end{minipage} \\[10pt]
\begin{minipage}[t]{0.4\linewidth}
\centering
\includegraphics[width=8cm]{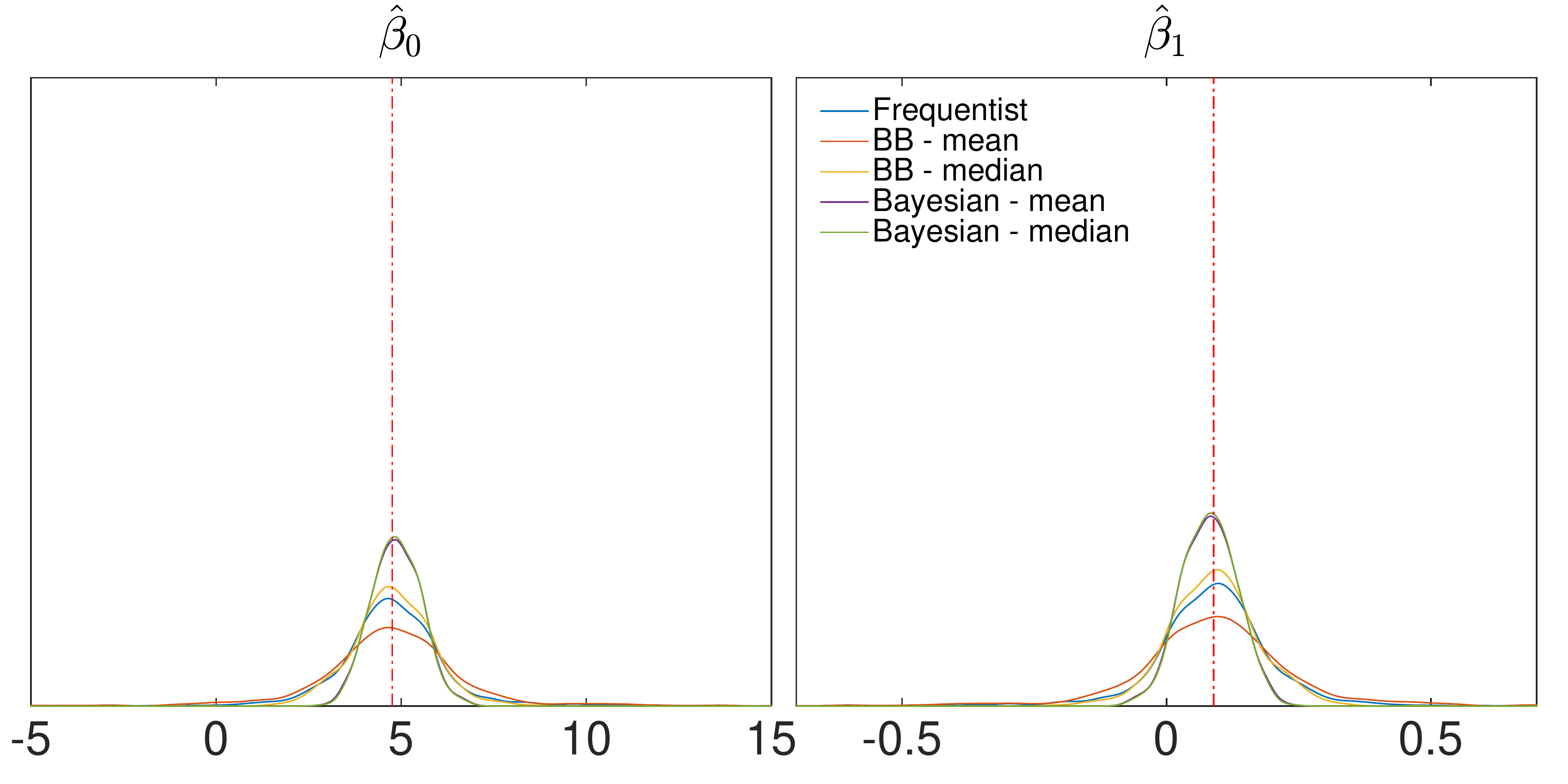} %
{\footnotesize $J=10,000$}
\end{minipage}
\hspace{2.0cm} 
\begin{minipage}[t]{0.4\linewidth}
\centering
\includegraphics[width=8cm]{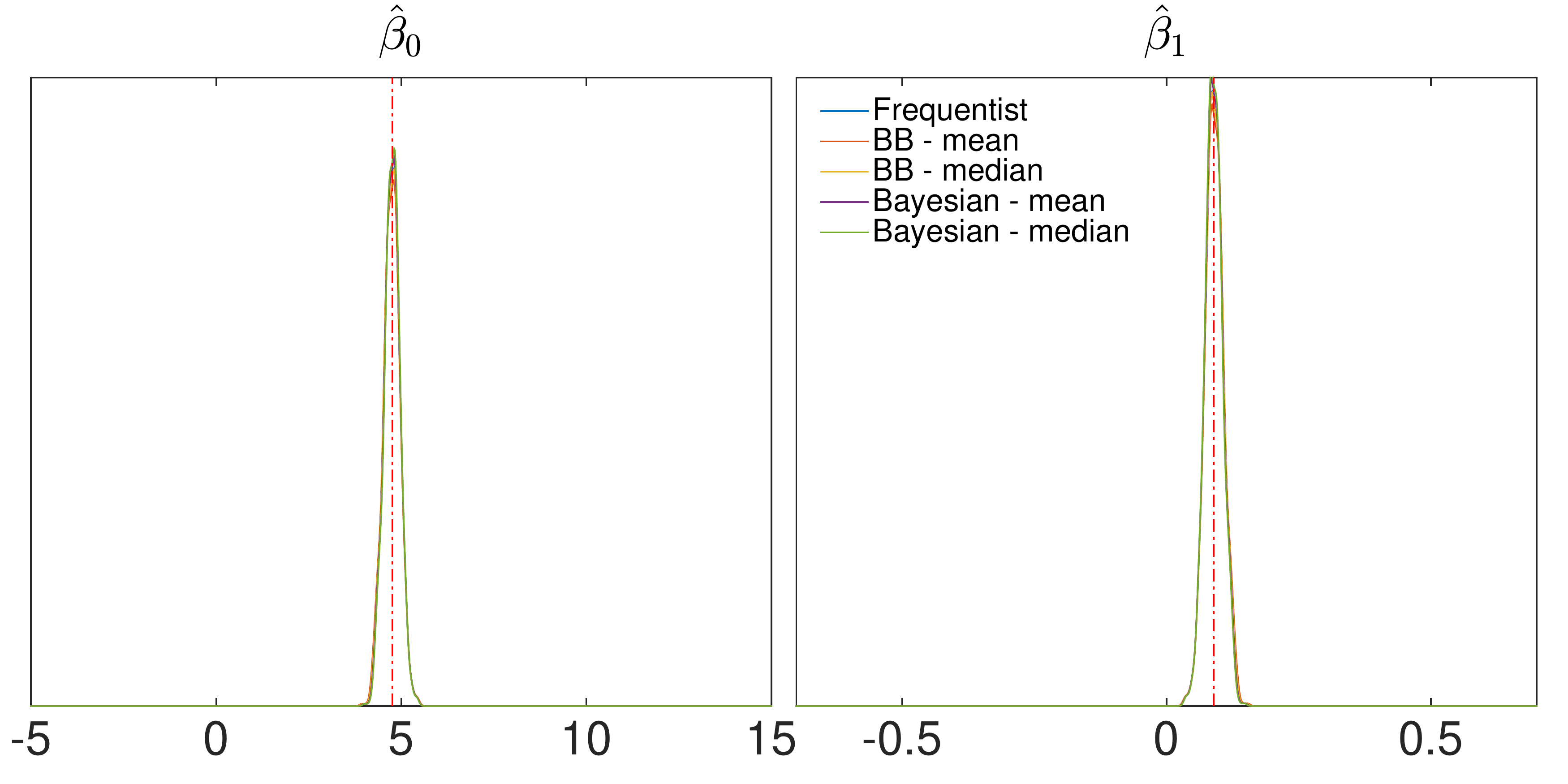} %
{\footnotesize $J=100,000$ }
\end{minipage}
\caption{The sampling distribution of classic IV (denoted frequentist),
Bayesian bootstrapping and Bayesian estimators of $\protect\beta $ in the
linear regression model with the instrumental variable employing sample
sizes $J=10$, $J=1,000$, $J=10,000$ and $J=100,000$.}
\label{fig:Example_IV_DistributionOfEstimators1}
\end{figure}

For $1,000$ iterations, a random sample of size $J$ has been drawn with
replacement from the $162,512$ population. For each replication the
resulting marginal prior distributions of $\beta _{0}$ and $\beta _{1}$
depend on the draws which generate the support and so vary over the $1,000 $
samples. Figure \ref{fig:Example_IV_MarginalPrior1} shows the pointwise $%
95\% $ confidence intervals of the marginal prior distributions over these $%
1,000$ replications, for $J=100$, $1,000$, $5,000$ and $10,000$. It shows
the information content of the prior is modest and only mildly depends upon
the random support and $J$, with less variation across replications in the
prior density as $J$ increases. \ Similar results have been obtained for
other sample sizes $J$.

For each random sample, we compute the classic IV estimates of $\beta $ and
the \cite{ChamberlainImbens(03)} Bayesian bootstrapping estimates obtained
by $10,000$ draws. For the latter we report both the means and the medians
as the estimators. These estimators are compared with the weakly informative
Bayesian estimators (using the prior described earlier).

The Bayesian estimates are obtained by the following resampling method.
Initially a sample of size $10,000$ is drawn from a Dirichlet distribution
with parameter $(n_{1}+\alpha -1,...,n_{J}+\alpha -1)$, and the importance
sampling weights are computed $w^{(k)}\propto \eta (\beta ^{(k)})$. Then a
sample from the posterior can be obtained by resampling using the normalized
weights. Estimators of the mean and the median of the posterior have been
reported here. \ For $J=10$, $100$, $1,000$, $5,000$, $10,000$, $40,000$ and 
$100,000$ the effective sample size divided by $J$ \cite[p. 35]{Liu(01)} was 
$0.620$, $0.576$, $0.607$, $0.719$, $0.819$, $0.978$ and $0.997$,
respectively. \ This suggests this is a reasonable method for this problem.
\ \ \ 

In Figure \ref{fig:Example_IV_DistributionOfEstimators1} the sampling
distribution of these five estimators have been plotted. The blue curves
correspond to the classical IV estimator. They exhibit a very imprecise
estimator and assign significant probabilities to economically irrelevant
values of $\beta $ (this is a well known disappointing property of this
estimator, e.g. \cite{BoundJaegerBaker(95)}). The mean of the Bayesian
bootstrapping estimator of \cite{ChamberlainImbens(03)} has a very large
variance too (the orange curves), but its median is more precise (the yellow
curves). The Bayesian estimators (that are the mean and the median of the
posterior) are the most precise estimators. 
\begin{table}[tbp]
\begin{center}
{\footnotesize 
\begin{tabular}{l|rrr|rrr|rrr|rrr}
\hline
& \multicolumn{12}{c}{$\beta_0$} \\ \hline
& \multicolumn{3}{|c}{Bias of mean} & \multicolumn{3}{|c}{Bias of median} & 
\multicolumn{3}{|c}{RMSE} & \multicolumn{3}{|c}{95\% CR length} \\ 
Sample size $J$ & 10 & 1,000 & 10,000 & 10 & 1,000 & 10,000 & 10 & 1,000 & 
10,000 & 10 & 1,000 & 10,000 \\ \hline\hline
Classic IV & -0.104 & 0.214 & -0.015 & 0.285 & 0.144 & -0.009 & 14.27 & 35.41
& 0.216 & 42.83 & 44.52 & 0.832 \\ 
BB $E(\theta|Z)$ & -0.174 & 0.909 & -0.020 & 0.347 & 0.259 & -0.015 & 50.01
& 27.32 & 0.221 & 42.21 & 45.36 & 0.851 \\ 
BB $med(\theta|Z)$ & 0.240 & 0.190 & -0.015 & 0.287 & 0.227 & -0.009 & 2.369
& 1.247 & 0.216 & 9.491 & 5.137 & 0.834 \\ 
$E(\theta|Z)$ & 0.323 & 0.269 & -0.007 & 0.324 & 0.292 & -0.003 & 0.979 & 
0.640 & 0.211 & 3.667 & 2.447 & 0.815 \\ 
$med(\theta|Z)$ & 0.324 & 0.261 & -0.002 & 0.326 & 0.290 & 0.003 & 1.034 & 
0.669 & 0.207 & 3.837 & 2.572 & 0.803 \\ \hline
& \multicolumn{12}{c}{$\beta_1$} \\ \hline
Classic IV & 0.007 & -0.016 & 0.001 & -0.017 & -0.011 & 0.001 & 1.100 & 2.783
& 0.017 & 3.398 & 3.496 & 0.065 \\ 
BB $E(\theta|Z)$ & -0.001 & -0.072 & 0.002 & -0.019 & -0.020 & 0.001 & 3.940
& 2.151 & 0.017 & 3.402 & 3.546 & 0.067 \\ 
BB $med(\theta|Z)$ & -0.017 & -0.015 & 0.001 & -0.018 & -0.017 & 0.001 & 
0.186 & 0.098 & 0.017 & 0.725 & 0.404 & 0.066 \\ 
$E(\theta|Z)$ & -0.023 & -0.021 & 0.001 & -0.020 & -0.022 & 0.000 & 0.077 & 
0.050 & 0.017 & 0.295 & 0.193 & 0.064 \\ 
$med(\theta|Z)$ & -0.023 & -0.020 & 0.000 & -0.021 & -0.022 & 0.000 & 0.081
& 0.052 & 0.016 & 0.307 & 0.200 & 0.063 \\ \hline
\end{tabular}%
}
\end{center}
\caption{Results for the linear regression with an instrument using $1,000$
replications sampling with replacement. The bias of the mean is the
difference of the mean of the replications and the true value (using all
162,512 data points). The bias of the median is the median of the
replications minus the true value. The 95\% confidence region (CR) length is
the length of 95\% of the replications placing 2.5\% of the mass in each
tail. RMSE is the root mean square error over the replications. BB denotes
the (non-informative) Bayesian bootstrap. med denotes median. The last two
rows are posterior mean and posterior median of the Bayesian model with
weakly informative prior.}
\label{Example_IV_RMSE1}
\end{table}

The bias (with its standard error) and the root mean square error (RMSE) of
the estimators have been reported in Table \ref{Example_IV_RMSE1}. Although
the Bayesian estimators are slightly biased, thanks to their small variances
they have lower RMSEs. In the Table and Figure \ref%
{fig:Example_IV_95CI_Length} we have also reported the length of the $95\%$
confidence intervals of the sampling distribution of the estimators (over
the 1,000 replications) of $\beta _{0}$ and $\beta _{1}$ for different
sample sizes $J=10,100,1,000,5,000,10,000,40,000$ and $100,000$. \ This
shows that the Bayesian estimators are far more accurate than the classical
IV estimator and Bayesian bootstrapping for most sample sizes. \ However,
when $J$ hits around $100,000$ the old methods catchup to our techniques. \
\ 

Why does our method do better? For weakly identified models even a very
modestly informative prior, which downweights economically implausible
values of the parameter space, has the trait of cutting off the tails of the
posterior corresponding to these implausible values. Because of the
ridge-like posterior induced by the weakly informative likelihood, the
posterior contracts onto a manifold, rather than a single point. As such,
having a prior which constrains the feasible support provides significant
value. 
\begin{figure}[tbp]
\begin{center}
\includegraphics[width=16cm]{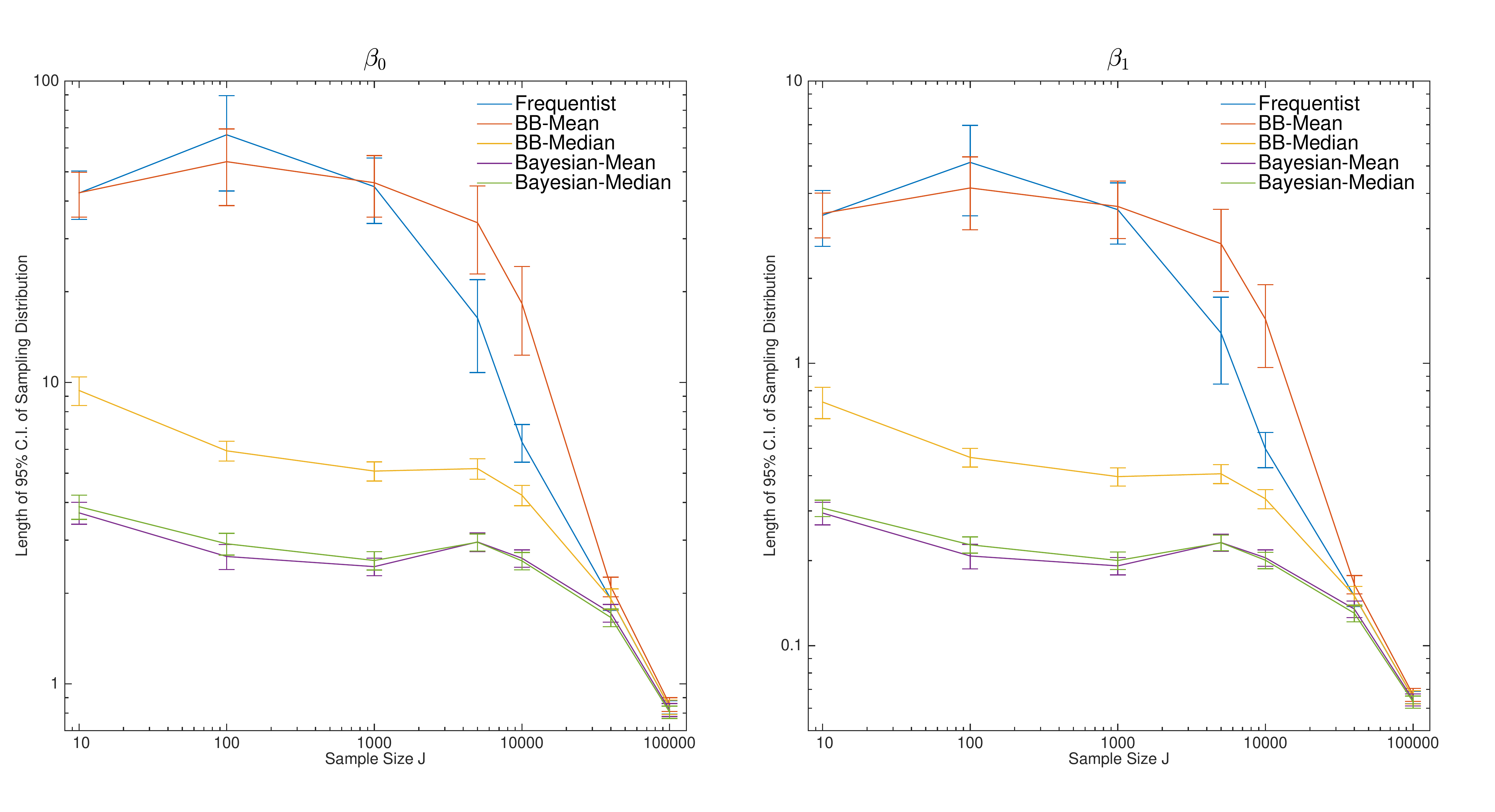}
\end{center}
\caption{The length of the $95\%$ confidence intervals of the sampling
distribution of the parameters $\protect\beta _{0}$ and $\protect\beta _{1}$
for different sample sizes $J=10,100,1,000,5,000,10,000,40,000$ and $100,000$%
, and for classical IV estimator, Bayesian bootstrapping (mean and median)
and Bayesian (mean and median). \ The bars denote our estimated 95\%
confidence intervals estimates of the lengths. \ }
\label{fig:Example_IV_95CI_Length}
\end{figure}

In the Appendix \ref{sect:app more support} we have relaxed the assumption
that the support of $Z$ is fully observed in our sample. It can be seen that
the estimates would not change significantly as long as $\alpha $, the
parameter of the Dirichlet distribution in the prior of $\theta ^{\ast }$,
is small. It can be shown that, when $\alpha \rightarrow 0^{+}$, the
marginal posterior distribution of $\theta $ and $\beta $ of both models
coincide.\newline

\subsection{Causal Inference \label{sect:empirical_studies_CausaIInference}}

In this example we analyze a dataset originally collected and studied in 
\cite{ImbensRubinSacerdote(01)}. The dataset contains socioeconomic
variables of $496$ individuals who had won monetary prizes in the
Massachusetts lottery. Following \cite{ImbensRubin(15)}, we call the
individuals who won large sums of money \textquotedblleft the winners" ($237$
observations), and the ones who won only small amounts \textquotedblleft the
losers" ($259$ observations). The goal is to study the effect of unearned
income on the economic behavior of the subjects, more specifically, on their
average labor income over the first six years following the year in which
they had won the lottery. For each individual the treatment indicator, $%
W_{i} $, is equal to one for the winners and zero for the losers. The
uncontroversial assumption behind this study is the random treatment
assignment, however one may argue that the sample is not representative of
the population. For instance in the literature it is well documented that
the lottery players are slightly more likely to be male and middle-aged,
with lower income and less education (see \cite{ClotfelterCook(89)}, \cite%
{FarrelWalker(99)} and \cite{Ariyabuddhiphongs(11)}, among others).

The dataset includes the year in which the winning lottery ticket is
purchased (\texttt{YW}), the number of tickets purchased in a typical week (%
\texttt{TB}), the individual's age (\texttt{Age}), gender (\texttt{G}) and
years of schooling (\texttt{YS}), an indicator showing whether she has been
working during the year the winning ticket is purchased (\texttt{WT}), and
the annual social security earnings from $6$ years prior to the year in
which the winning ticket is purchased (\texttt{EYB1} to \texttt{EYB6}) to $6$
years after that (\texttt{EYA1} to \texttt{EYA6}), all converted to $1986$
dollars. The authors argue, perhaps optimistically, that the social security
income is potentially the most reliable measure of income in long run,
although it is capped to the maximum taxable earning ($\$42,000$ in $1986$).

In order to improve the overlap of the background variables, following the
recommendation of \cite{ImbensRubin(15)}, initially we model the propensity
scores using a logistic regression model, and estimate the model's parameter
using the Bayesian bootstrapping of \cite{ChamberlainImbens(03)}. The
covariates of the model are a constant, the linear terms \texttt{TB}, 
\texttt{YS}, \texttt{WT}, \texttt{EYB1}, \texttt{Age}, \texttt{YW}, the
indicator for the positiveness of the earning $5$ years before winning the
lottery (\texttt{SEYB5}), \texttt{G}, and the quadratic terms \texttt{YW} $%
\times $ \texttt{YW}, \texttt{EYB1} $\times $ \texttt{G}, \texttt{TB} $%
\times $ \texttt{TB}, \texttt{TB} $\times $ \texttt{WT}, \texttt{YS} $\times 
$ \texttt{YS}, \texttt{YS} $\times $ \texttt{EYB1}, \texttt{TB} $\times $ 
\texttt{YS}, \texttt{EYB1} $\times $ \texttt{Age}, \texttt{Age} $\times $ 
\texttt{Age}, and \texttt{YW} $\times $ \texttt{G}. We discard the
observations with too small ($<0.0891$) or too large ($>0.9109$) estimates
of propensity scores. This results in a sample of size $N=295$ ($142$
winners and $153$ losers). In the proposed model the propensity score is
regressed on $13$ covariates using a logistic regression. The vector of
covariates is denoted by $X_{i}$, and include a constant, the linear terms 
\texttt{TB}, \texttt{YS}, \texttt{WT}, \texttt{EYB1}, \texttt{Age}, \texttt{%
SEYB5}, \texttt{YW}, \texttt{EYB5}, and the quadratic terms \texttt{YW} $%
\times $ \texttt{YW}, \texttt{TB} $\times $ \texttt{YW}, \texttt{TB} $\times 
$ \texttt{TB}, and \texttt{WT} $\times $ \texttt{YW}. For details on the
variable selection see \cite{ImbensRubin(15)}. The outcome, $Y_{i}$, is the
average of the individual's income averaged over the first $6$ years after
purchasing the winning lottery ticket. Therefore the parameters of the
logistic regression model, $\gamma $, and the ATE, $\tau $, satisfy the
following moment conditions, 
\begin{equation}
\mathbb{E}\left[ g(Z_{i},\beta )\right] =0,
\end{equation}%
in which, $Z_{i}=(X_{i},Y_{i},W_{i})$, $\beta =(\gamma ,\tau )$, and, 
\begin{equation}
g(Z_{i},\beta )=\left[ 
\begin{array}{c}
X_{i}(Y_{i}-\eta _{i}) \\ 
\frac{(W_{i}-\eta _{i})Y_{i}}{\eta _{i}\left( 1-\eta _{i}\right) }-\tau \\ 
\end{array}%
\right] ,
\end{equation}%
where $\eta _{i}=\frac{\exp (\gamma ^{\prime }X_{i})}{1+\exp (\gamma
^{\prime }X_{i})}$. If we assume $Z_{i}$s are i.i.d. draws from a discrete
distribution supported on $\{s_{1},...,s_{J}\}$, with $\mathbb{P}%
(Z_{i}=s_{j})=\theta _{j}$, the parameters $(\beta ,\theta )$ will satisfy
the following system of equations, 
\begin{equation}
\left[ 
\begin{array}{c}
\sum_{j=1}^{J}\theta _{j}x_{j}(y_{j}-\eta _{j}) \\ 
\sum_{j=1}^{J}\theta _{j}\frac{(w_{j}-\eta _{j})y_{j}}{\eta _{j}\left(
1-\eta _{j}\right) }-\tau \\ 
\end{array}%
\right] =0.
\end{equation}%
We let the prior of $(\beta ,\theta )$ be 
\begin{equation}
p(\beta ,\theta )\propto \frac{1}{\sqrt{J_{\theta }J_{\theta }^{\prime
}+I_{14}}}\eta (\gamma )\eta (\tau )\eta (\theta )1_{\Theta _{\beta ,\theta
}}(\beta ,\theta ),
\end{equation}%
in which the initial prior of the regression coefficients, $\eta (\gamma )$,
is a normal distribution centered at their estimates obtained from the
Bayesian bootstrap of \cite{ChamberlainImbens(03)} and its covariance matrix
is equal to the covariance matrix of estimates scaled by a factor of $100$,
and the initial prior of ATE is a zero mean normal distribution with
variance equal to $100$. Moreover we use a symmetric Dirichlet distribution
with parameter $\alpha =10^{-6}$ as the initial prior on $\theta $.

\begin{figure}[tbp]
\begin{center}
\includegraphics[width=12cm]{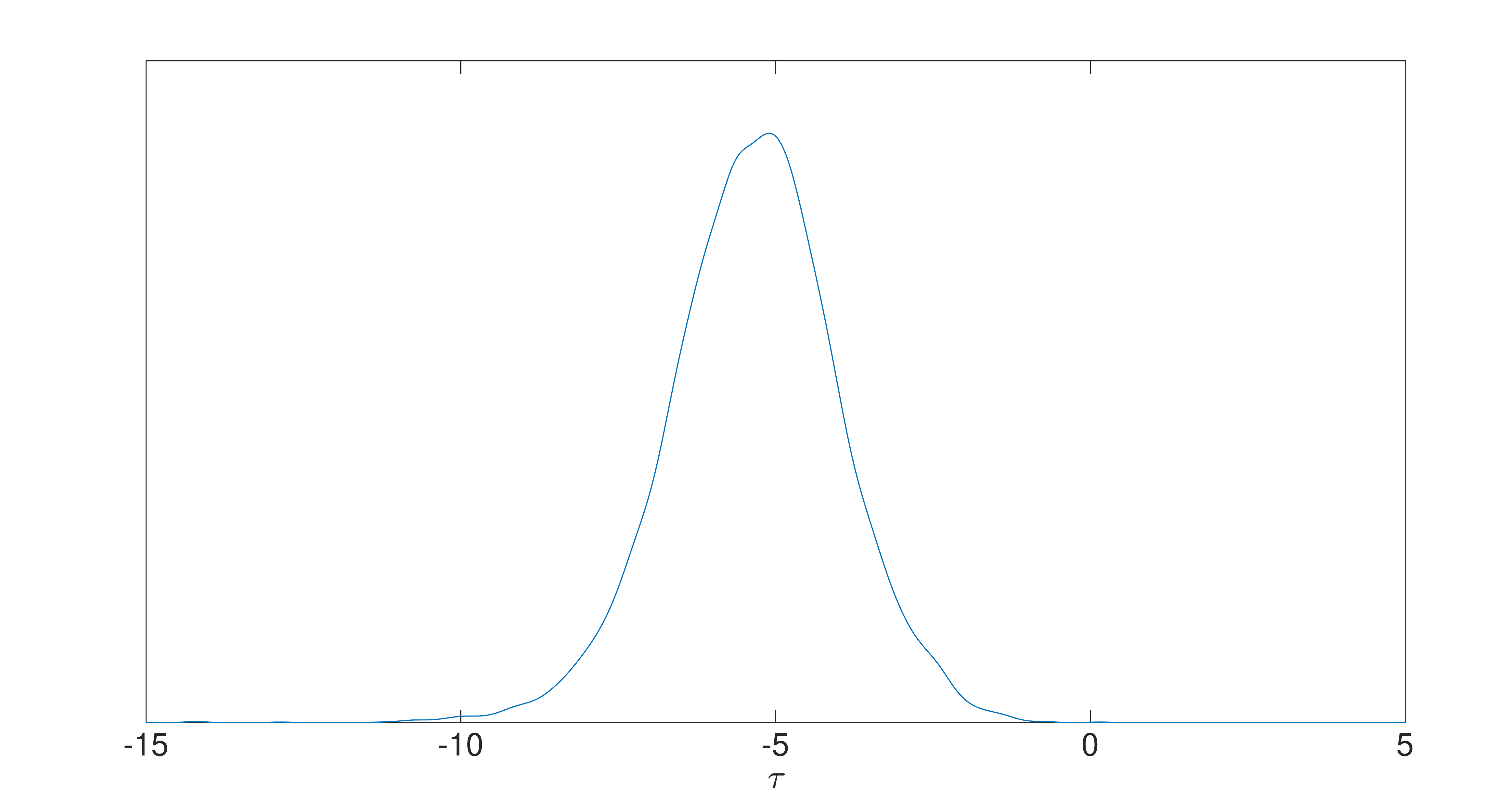}
\end{center}
\caption{The posterior distribution of the average treatment effect (ATE) on
subsequent annual earnings of a substantial lottery win for the lottery data
set.}
\label{fig:ATE_Posterior}
\end{figure}

By reweighting draws from the posterior distribution of the Bayesian
bootstrap of \cite{ChamberlainImbens(03)}, we obtain $10,000$ independent
draws from the posterior of our model. An estimate of the posterior
distribution of the ATE is depicted in Figure \ref{fig:ATE_Posterior}. 
\textit{A posteriori} the expected value of ATE is $-\$5,346$ (with $95\%$
credible interval of $[-\$8,069,-\$2,720]$). This indicates that the average
income of the winners of the lotteries, in the years after winning the
prize, tend to slightly decrease. Our estimate of ATE is only slightly
different from the frequentist estimate.

\section{Conclusions\label{section:conclusions}}

In this paper we have provided a coherent Bayesian calculus for rational
nonparametric moment based estimators, allowing users to specify
scientifically meaningful priors. At the core of our analysis is a prior
density placed on the Hausdorff measure whose support is generated by the
scientific parameters of interest and the nonparametric probabilities. We
show how to transform this prior into a posterior density.

Much moment based analysis\ favoured in the literature delivers weakly
identified parameters. \ The use of very modest priors can dramatically
improve estimation by downweighting vast regions of economically implausible
parameter values. \ Such weak priors play little role when the data is
informative but provide a safety net when this is not the case. \ \ 

To harness these gains, at the center of our paper are the marginal method
and the joint method. The first is based on finding the density of the
probabilities with respect to a Lebesgue measure. This allows for the use of
conventional simulation methods such as MCMC, importance sampling and
Hamiltonian Monte Carlo. It is convenient to use where the moment conditions
can be solved analytically or numerically very fast. \ 

Our joint method is somewhat harder to code but has the virtue of never
having to solve the moment equations. This has some speed advantages but
more fundamentally allows the rational analysis of moment condition models
with many solutions. As a side product our method provides a novel way of
generically simulating on a wide class of manifolds, which may be useful in
other areas of science. \ \ \ 

\baselineskip=14pt

\bibliographystyle{chicago}
\bibliography{REZA}

\baselineskip=20pt

\appendix%

\section{Appendices}

\subsection{Proof of proposition \protect\ref{Prop:marginal}}

Since corresponding to every $\theta \in \Theta _{\theta }$ there is a
unique $\beta $, there exist a one-to-one mapping between $\Theta _{\beta
,\theta }$ and $\Theta _{\theta }$: $(\beta ,\theta )=\left\{ \beta (\theta
),\theta )\right\} =F(\theta )$. Now let $A$ be a measurable set on $\Theta
_{\beta ,\theta }$, and assume $S_{\theta }(A)$ is its projection on $\Theta
_{\theta }$. Therefore 
\begin{equation}
\mathbb{P}(S_{\theta }(A))=\mathbb{P}(A)=\int_{A}p(\beta ,\theta
)dA=\int_{S_{\theta }(A)}\left\Vert v_{1}\wedge \cdots \wedge
v_{J-1}\right\Vert p(\beta ,\theta )dS  \notag
\end{equation}%
where $v_{j}=\frac{\partial F}{\partial \theta _{j}}$ (for $1\leq j\leq J-1$%
). Therefore $\left\Vert v_{1}\wedge \cdots \wedge v_{J-1}\right\Vert
p(\beta ,\theta )$ is the density of $\theta $ with respect to Lebesgue
measure. Moreover, 
\begin{equation*}
\left\Vert v_{1}\wedge \cdots \wedge v_{J-1}\right\Vert =\left[
Gram(v_{1},...,v_{J-1})\right] ^{\frac{1}{2}}=\left\vert J_{\theta }^{\prime
}J_{\theta }+I_{J-1}\right\vert ^{\frac{1}{2}}=\left\vert J_{\theta
}J_{\theta }^{\prime }+I_{p}\right\vert ^{\frac{1}{2}}
\end{equation*}%
where $Gram(\cdot )$ is the Gramian determinant and $J_{\theta }=\partial
\beta /\partial \theta ^{^{\prime }}$.\newline

\subsection{Proof of proposition \protect\ref{prop:joint method}}

Let $p(\beta )$ be the density of $\beta $. Then, given $\beta $, the vector
of probabilities $\theta $ lives on a $J-1-p$ dimensional hyperplane in $%
\mathbb{R}^{J-1}$ defined by $H\theta +g_{J}=0$. This system of equation can
be solved for $p$ elements of the variables $\theta
_{J-p:J-1}=-H_{2}^{-1}\left( H_{1}\theta _{1:J-p-1}-g_{J}\right) $, where $%
H_{1}=[h_{1}\ ...\ h_{J-p-1}]$ and $H_{2}=[h_{J-p}\ ...\ h_{J-1}]$.
Therefore, $\partial \theta _{J-p:J-1}/\partial \theta
_{1:J-p-1}=-H_{2}^{-1}H_{1}$ and so 
\begin{eqnarray*}
p(\theta _{1:J-p-1}|\beta ) &=&\left\vert H_{2}^{-1}H_{1}H_{1}^{\prime
}H_{2}^{^{\prime }-1}+I_{p}\right\vert ^{\frac{1}{2}}p(\theta |\beta ), \\
p(\theta _{1:J-p-1},\beta ) &=&\left\vert H_{2}^{-1}H_{1}H_{1}^{\prime
}H_{2}^{^{\prime }-1}+I_{p}\right\vert ^{\frac{1}{2}}p(\beta )p(\theta
|\beta ).
\end{eqnarray*}%
Therefore the density of $\theta $ is 
\begin{eqnarray}
p(\theta ) &=&\left\vert \frac{\partial (\theta _{1:J-p-1},\beta )}{\partial
(\theta )}\right\vert p(\theta _{1:J-p-1},\beta )=\left\vert \frac{\partial
\beta }{\partial \theta _{J-p:J-1}}\right\vert p(\theta _{1:J-p-1},\beta ) 
\notag \\
&=&\left\vert \left[ \mathbb{E}\left( \frac{\partial g}{\partial \beta
^{\prime }}\right) \right] ^{-1}H_{2}\right\vert p(\theta _{1:J-p-1},\beta )
\notag \\
&=&\left\vert \left[ \mathbb{E}\left( \frac{\partial g}{\partial \beta
^{\prime }}\right) \right] ^{-1}H_{2}\right\vert \left\vert
H_{2}^{-1}H_{1}H_{1}^{\prime }H_{2}^{^{\prime }-1}+I_{p}\right\vert ^{\frac{1%
}{2}}p(\beta )p(\theta |\beta )  \notag \\
&=&\left\vert \left[ \mathbb{E}\left( \frac{\partial g}{\partial \beta
^{\prime }}\right) \right] ^{-1}\right\vert \left\vert H_{1}H_{1}^{\prime
}+H_{2}H_{2}^{\prime }\right\vert ^{\frac{1}{2}}p(\beta )p(\theta |\beta
)=\left\vert \left[ \mathbb{E}\left( \frac{\partial g}{\partial \beta
^{\prime }}\right) \right] ^{-1}\right\vert \left\vert HH^{\prime
}\right\vert ^{\frac{1}{2}}p(\beta )p(\theta |\beta )  \notag \\
&=&\left\vert \left[ \mathbb{E}\left( \frac{\partial g}{\partial \beta
^{\prime }}\right) \right] ^{-1}HH^{\prime }\left[ \mathbb{E}\left( \frac{%
\partial g^{\prime }}{\partial \beta }\right) \right] ^{-1}\right\vert ^{%
\frac{1}{2}}p(\beta )p(\theta |\beta )=\left\vert \frac{\partial \beta }{%
\partial \theta ^{^{\prime }}}\frac{\partial \beta ^{\prime }}{\partial
\theta }\right\vert ^{\frac{1}{2}}p(\beta )p(\theta |\beta ).  \notag
\end{eqnarray}%
Therefore: 
\begin{equation*}
p(\beta ,\theta )=\frac{\left\vert \frac{\partial \beta }{\partial \theta
^{^{\prime }}}\frac{\partial \beta ^{\prime }}{\partial \theta }\right\vert
^{\frac{1}{2}}}{\left\vert \frac{\partial \beta }{\partial \theta ^{^{\prime
}}}\frac{\partial \beta ^{\prime }}{\partial \theta }+I_{p}\right\vert ^{%
\frac{1}{2}}}p(\beta )p(\theta |\beta ).
\end{equation*}

\subsection{Joint method proposal\label{sect:joint method proposal}}

In order to generate a proposal value for $\theta^*$, we can first draw $\pi
^{\ast }$ from $\mathcal{N}(\theta ,\Sigma _{Q})$, and let $\theta ^{\ast }$
be the closest point to $\pi ^{\ast }$ in the hyperplane $\mathcal{P}^* =
\{\lambda \in \mathbb{R}^{J-1};H^{\ast }\lambda +g_{J}^{\ast }=0\}$, where
we measure the distance between $\pi ^{\ast }$ and $\theta ^{\ast }$ with
the squared Euclidean norm: 
\begin{equation*}
\theta ^{\ast }=\underset{\theta }{\func{argmin}}\ \frac{1}{2}\left\Vert \pi
^{\ast }-\theta \right\Vert _{2}^{2}+\frac{1}{2}\left( \iota ^{\prime }\pi
^{\ast }-\iota ^{\prime }\theta \right) ^{2}.
\end{equation*}%
The quadratic penalty is certainly inelegant (e.g. compared to the
log-likelihood of the multinomial model, but see, for example, \cite%
{Owen(91)} and \cite{AntoineBonnalRenault(07)} who use it for their
Euclidean empirical likelihood) as the resulting $\theta ^{\ast }$ can have
negative elements or may result in $\theta _{J}^{\ast }=1-\iota ^{\prime
}\theta ^{\ast }\leq 0$. However, by using a quadratic penalty, $\theta
^{\ast }$ becomes the solution to a quadratic optimization problem subject
to $p$ equality constraints, and so has an analytic solution $\theta ^{\ast
}=a^{\ast }+B^{\ast }\pi ^{\ast }$.

The Lagrangian of the optimization is, 
\begin{equation*}
E(\theta ,\lambda )=\left\Vert \pi ^{\ast }-\theta \right\Vert
_{2}^{2}+\left( \iota ^{\prime }\pi ^{\ast }-\iota ^{\prime }\theta \right)
^{2}+\lambda ^{\prime \ast }\theta +g_{J}^{\ast })
\end{equation*}%
and the first order conditions are: 
\begin{equation}
\frac{\partial E}{\partial \theta }=(I+\iota \iota ^{\prime \ast }-\pi
^{\ast })+H^{\ast ^{\prime }}\lambda =0,\quad \frac{\partial E}{\partial
\lambda }=H^{\ast }\theta ^{\ast }+g_{J}^{\ast }=0.  \notag
\end{equation}%
Solving them for $\theta ^{\ast }$ and $\lambda $ results in, 
\begin{eqnarray}
\theta ^{\ast } &=&\pi ^{\ast }-(I+\iota \iota ^{\prime -1}H^{\ast ^{\prime
}}\left[ H^{\ast }(I+\iota \iota ^{\prime -1}H^{\ast ^{\prime }}\right]
^{-1}(H^{\ast }\pi ^{\ast }+g_{J}^{\ast })  \notag \\
\lambda &=&\left[ H^{\ast }(I+\iota \iota ^{\prime -1}H^{\ast ^{\prime }}%
\right] ^{-1}(H^{\ast }\pi ^{\ast }+g_{J}^{\ast }).  \notag
\end{eqnarray}%
Therefore $\theta ^{\ast }$ is an affine transformation of $\pi ^{\ast }$: $%
\theta ^{\ast }=a^{\ast }+B^{\ast }\pi ^{\ast }$, where 
\begin{eqnarray}
a^{\ast } &=&-(I+\iota \iota ^{\prime -1}H^{\ast ^{\prime }}\left[ H^{\ast
}(I+\iota \iota ^{\prime -1}H^{\ast ^{\prime }}\right] ^{-1}g_{J}^{\ast } 
\notag \\
B^{\ast } &=&I-(I+\iota \iota ^{\prime -1}H^{\ast ^{\prime }}\left[ H^{\ast
}(I+\iota \iota ^{\prime -1}H^{\ast ^{\prime }}\right] ^{-1}H^{\ast }. 
\notag
\end{eqnarray}

This transformation from $\pi ^{\ast }$ to $\theta ^{\ast }$ is a
many-to-one affine transformation. Consequently, $\theta ^{\ast }|\beta
^{\ast },\beta ^{(t)},\theta ^{(t)}$ is a singular normal distribution with
mean $a^{\ast }+B^{\ast }\theta ^{(t)}$ and variance matrix $B^{\ast }\Sigma
_{Q}B^{\ast }$.

A singular normal distribution with mean $\mu $ and (singular) variance
matrix $\Sigma $ has a density on the range of the covariance matrix (e.g. 
\cite{Khatri(68)}), given by 
\begin{equation*}
(2\pi )^{-\frac{1}{2}rank(\Sigma )}|\Sigma |_{\QTR{md}{rank}(\Sigma )}^{-%
\frac{1}{2}}\exp \left\{ -\frac{1}{2}\left( x-\mu \right) ^{\prime }\Sigma
^{+}\left( x-\mu \right) \right\} ,
\end{equation*}%
where $|\Sigma |_{\QTR{md}{rank}(\Sigma )}$ is the product of non-zero
eigenvalues of $\Sigma $ and $\Sigma ^{+}$ is its Moore-Penrose inverse.

In our algorithm, $\Sigma _{Q}$ and the parameters inside $q(\cdot |\beta
^{(t)},\theta ^{(t)})$ are the tuning parameters. We may either adapt them
in the course of simulation, or they can be set to some fixed values
obtained from an estimate of the posterior's distribution. Here we document
how we have carried this out for our simulation and empirical work. A simple
to calculate candidate for the covariance of $\beta $'s proposal is $\Sigma
_{\beta }=\left( \Sigma _{0_{\mathbf{\beta }}}^{-1}+\Sigma _{\QTR{md}{BB}_{%
\mathbf{\beta }}}^{-1}\right) ^{-1}$, where $\Sigma _{0_{\mathbf{\beta }%
}}^{-1}$ is the prior's covariance and $\Sigma _{\QTR{md}{BB}_{\mathbf{\beta 
}}}^{-1}$ is the covariance of the estimates of $\beta $ obtained by
Bayesian bootstrapping of \cite{ChamberlainImbens(03)} (As an alternative we
may use the asymptotic covariance of the least squares or GMM estimators).
Moreover a suitable candidate for $\Sigma _{Q}$ is $diag(\hat{\theta}%
_{1}^{2},...,\hat{\theta}_{J-1}^{2})$ where: 
\begin{equation}
\hat{\theta}=(\hat{\theta}_{1},...,\hat{\theta}_{J-1})=\underset{\theta }{%
\func{argmax}}\sum_{j=1}^{J}n_{j}\ln \theta _{j}\ \ \ \text{subject to}\ \ \ 
\hat{H}\theta +\hat{g}_{J}=0,
\end{equation}%
in which $\hat{H}=\left( \hat{g}_{1},...,\hat{g}_{J-1}\right) -\hat{g}%
_{J}\iota ^{\prime }$, $\hat{g}_{j}=g(\hat{\beta},s_{j})$ and $\hat{{\beta }}%
=\left( {\Sigma }_{0_{{\beta }}}+{\Sigma }_{\QTR{md}{BB}_{{\beta }}}\right)
^{-1}\left( {\Sigma }_{0_{{\beta }}}{\mu }_{\QTR{md}{BB}_{{\beta }}}+{\Sigma 
}_{\QTR{md}{BB}_{{\beta }}}{\mu }_{0_{{\beta }}}\right) $.

\subsection{Large support\label{sect:large support}}

An apparent drawback of the joint method is that in each evaluation of the
proposal's density, the Moore-Penrose inverse of the $(J-1)\times (J-1)$
matrix $B^{\ast }\Sigma _{Q}B^{\ast \prime }$ should be computed. In general
this costs $O(J^{3})$ computational operations. \ \ This type of challenge
is very common in Bayesian analysis and a standard approach to this problem
is to make proposals to update a block of $K\ll J$ elements of $\theta $,
with cost $O(K^{3})$. \ 

%

\begin{figure}[tbp]
\begin{center}
\includegraphics[width=17cm]{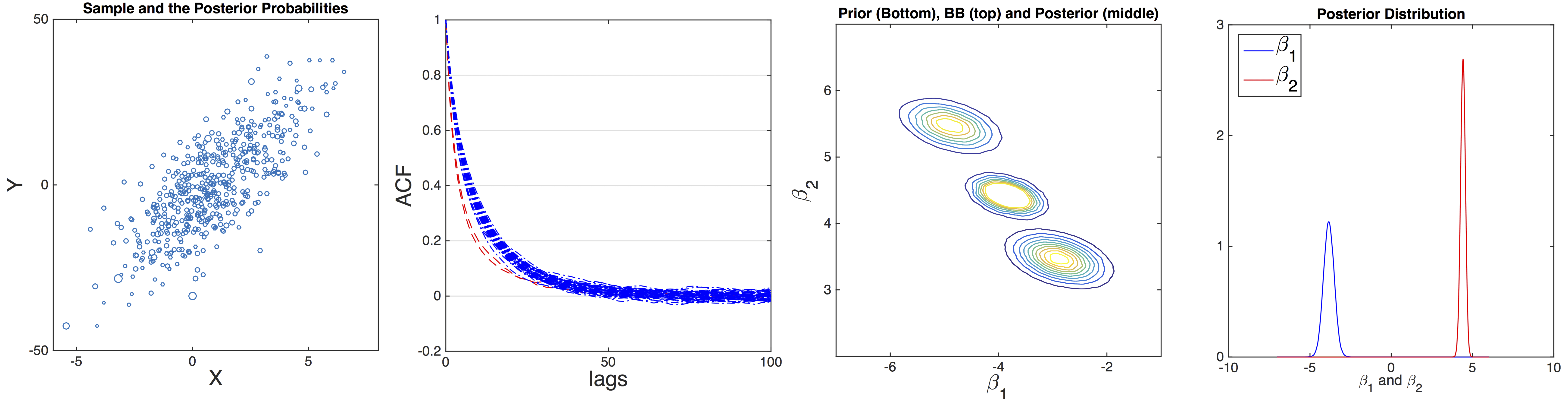}
\end{center}
\caption{{\protect\small Inference in linear regression model with }$J=500$%
{\protect\small \ and an informative prior.}}
\label{fig:Example_LRJ500_Plots_InformativePrior}
\end{figure}

Let the $K\times 1$ vector $u$ be a randomly (without replacement) selected
subset of the indices $\{1,...,J-1\}$ and the $(J-K-1)\times 1$ vector $v$
be its complement. Moreover let $\tilde{\theta}=(\theta _{u_{1}},...,\theta
_{u_{K}})$ and $\bar{\theta}=(\theta _{v_{1}},...,\theta _{v_{J-K-1}})$. The
proposal's vector of probabilities, $\theta ^{\ast }$, is equal to $\theta $
except for the $K$ elements with indices in $u$, $\tilde{\theta}^{\ast
}=(\theta _{u_{1}}^{\ast },...,\theta _{u_{K}}^{\ast })$, that is obtained
by solving: 
\begin{equation}
\tilde{\theta}^{\ast }=\underset{\tilde{\theta}}{\func{argmin}}\ \frac{1}{2}%
\Vert \tilde{\theta}-\tilde{\pi}^{\ast }\Vert +\frac{1}{2}\left( \iota
^{\prime }\tilde{\theta}-\iota ^{\prime }\tilde{\pi}^{\ast }\right) \ \ \ 
\text{subject to}\ \ \ \bar{H}^{\ast }\bar{\theta}^{(t)}+\tilde{H}^{\ast }%
\tilde{\theta}+g_{J}^{\ast }=0,
\end{equation}%
where $\tilde{H}^{\ast }=\left( g_{u_{1}}^{\ast },...,g_{u_{K}}^{\ast
}\right) -g_{J}^{\ast }\iota ^{\prime }$, $\bar{H}^{\ast }=\left(
g_{v_{1}}^{\ast },...,g_{v_{J-K-1}}^{\ast }\right) -g_{J}^{\ast }\iota
^{\prime }$, and $\tilde{\pi}^{\ast }$ is a random draw from $N(\tilde{\theta%
},\Sigma _{\tilde{\theta}})$. Again this is a quadratic optimization problem
subject to a set of equality constraints with the following solution: $%
\tilde{\theta}^{\ast }=\tilde{a}^{\ast }+\tilde{B}^{\ast }\tilde{\pi}^{\ast
} $, where 
\begin{eqnarray}
\tilde{a}^{\ast } &=&-(I+\iota \iota ^{\prime -1}\tilde{H}^{\ast ^{\prime }}%
\left[ \tilde{H}^{\ast }(I+\iota \iota ^{\prime -1}\tilde{H}^{\ast ^{\prime
}}\right] ^{-1}\left( \bar{H}^{\ast }\bar{\theta}^{(t)}+g_{J}^{\ast }\right)
\notag \\
\tilde{B}^{\ast } &=&I-(I+\iota \iota ^{\prime -1}\tilde{H}^{\ast ^{\prime }}%
\left[ \tilde{H}^{\ast }(I+\iota \iota ^{\prime -1}\tilde{H}^{\ast ^{\prime
}}\right] ^{-1}\tilde{H}^{\ast }.  \notag
\end{eqnarray}

\subsection{Linear regression with an informative prior\label{sect:appendix
linear regression}}

Here we report the results for the linear regression model with sample size $%
J=500$, and an informative prior for $\beta $. We place a normal prior on $%
\beta $ with the mean equal to $\hat{\beta}_{\text{MLE}}+(5,-5)^{\prime }$
and the variance equal to the asymptotic variance of $\hat{\beta}_{\text{MLE}%
}$. Therefore the prior is as informative as the data, however centered at a
significantly different point. \ 

Figure \ref{fig:Example_LRJ500_Plots_InformativePrior}'s top left panel
shows a scatter plot of the sample. Each circle represents a data point and
its radius is proportional to $\mathbb{E}(\theta _{j}|Z)$. In the top-right
the ACF of the chains of $\beta $ and $50$ elements of $\theta $ have been
presented (the red dashed lines and the blue dotted lines correspond to $%
\beta $ and $\theta $, respectively.) These show that the Markov chain is
mixing sufficiently well. In the bottom-left panel the contour plot of the
prior distribution (bottom), posterior distribution of $\beta $ using
Bayesian bootstrapping, and the posterior distribution of $\beta $
considering the informative prior (middle) have been depicted. In the
bottom-right panel the histogram of the samples from the posterior of $\beta 
$ can be seen.

\subsection{Instrumental variables with partially observed support\label%
{sect:app more support}}

Now we assume the support of $S=(X,Y,W)$ has other $J$ missing elements (not
observed in the data), therefore $J^{\ast }=2J$. The density of our prior
for the missing elements of the support is, 
\begin{equation}
f_{S}(\tilde{s})=f_{S^{(1)}}(\tilde{s}^{(1)})f_{S^{(2)}}(\tilde{s}%
^{(2)})f_{S^{(3)}}(\tilde{s}^{(3)})
\end{equation}%
in which $f_{S^{(1)}}(\tilde{s}^{(1)})$ and $f_{S^{(3)}}(\tilde{s}^{(3)})$
are the density of a uniform distribution on $\{0,1,...,20\}$ and $\{0,1\}$,
respectively, and $f_{S^{(2)}}(\tilde{s}^{(2)})$ is a normal density with
mean $6$ and standard deviation $3$. Moreover we assume, 
\begin{equation}
p(\beta ,\theta ^{\ast }|\tilde{S})\propto \frac{1}{\sqrt{J_{\theta ^{\ast
}}J_{\theta ^{\ast }}^{\prime }+J_{\tilde{S}}J_{\tilde{S}}^{\prime }+I_{2}}}%
\eta (\beta )\eta (\theta ^{\ast })1_{\Theta _{\beta ,\theta ^{\ast },\tilde{%
S}}}(\beta ,\theta ^{\ast },\tilde{S}),
\end{equation}%
where $\eta (\beta )=\varphi \left( \beta _{0};5,4\right) \varphi \left(
\beta _{1};0,0.2\right) $ (similar to the previous case), and $\eta (\theta
^{\ast })$ is the density a symmetric Dirichlet distribution with $\alpha
=10^{-6}$. Hence the posterior distribution of $(\theta ^{\ast },\tilde{S})$
is: 
\begin{equation}
p(\theta ^{\ast },\tilde{S}|Z)\propto \eta (\beta )\left( \prod_{j=1}^{%
\tilde{J}}f_{S}(\tilde{s}_{j})\right) \left( \prod_{j=1}^{J^{\ast }}\theta
_{j}^{\ast ^{n_{j}+\alpha -1}}\right) .
\end{equation}

To sample from this distribution we can reweight random draws from the
following proposal, 
\begin{equation}
g(\theta ^{\ast },\tilde{S}|Z)\propto \left( \prod_{j=1}^{\tilde{J}}f_{S}(%
\tilde{s}_{j})\right) \left( \prod_{j=1}^{J^{\ast }}\theta _{j}^{\ast
^{n_{j}+\alpha -1}}\right) ,
\end{equation}%
with he weights proportional to $\eta (\beta )$. Now we set $J=10$, and for $%
1000$ times we draw a random sample from our dataset. Then we compare the
posterior distribution of the parameters under two assumptions. In the first
model we assume the support of $S$ is fully observed in the data (similar to
the previous section of this example), while in the second model we assume
the data has $\tilde{J}=10$ more elements that are not observed in our
sample. Since the prior of the probabilities and $\beta $ is barely
informative the posterior distributions of $\beta $ are almost
indistinguishable under these two assumptions. 

\subsection{Not the just identified case\label{sect:not just identified}}

\subsubsection{Abstract expression of the problem}

Collect all the parameters in the model and constraints as 
\begin{equation*}
\psi =(\theta _{1},...,\theta _{J-1},\beta _{1},...,\beta _{p})^{\prime
},\quad g(\psi )=0_{r}.
\end{equation*}%
Then resulting constrained support is $\psi \in \Theta _{\psi }$. Write $%
\lambda =\psi _{\mathcal{I}}$, $\phi =\psi _{\mathcal{I}^{c}}$, where $%
\mathcal{I}$ selects distinct indexes of $\psi $ and $\mathcal{I}^{c}$ is
the complement, so $\mathcal{I\cup I}^{c}=\left\{ 1,2,...,p+J-1\right\} $. \
Throughout we take $\dim (\phi )=r$ and consequently $\dim (\lambda )=J-m$,
where $m=r-p+1$.

Given the freedom to build $\mathcal{I}$ we make the following assumption. \ 

\textbf{Assumption A.} \ \textit{Under }$g(\psi )=0$\textit{\ knowledge of }$%
\lambda $\textit{\ reveals }$\phi $\textit{, so there exists a unique }$\phi
=t(\lambda )$\textit{. }

\subsubsection{Marginal method}

Under Assumption A, the area formula implies that $p(\lambda )=p(\psi )\sqrt{%
\left\vert I_{r}+J_{\phi \lambda }J_{\phi \lambda }^{\prime }\right\vert }$, 
$J_{\phi \lambda }=\partial \phi /\partial \lambda ^{\prime }$, where $%
p(\psi )$ is a density with respect to the $(J-1+p-r)$-dimensional Hausdorff
measure on $\Theta _{\psi }$, while $p(\lambda )$ is a density with respect
to the $J-m$-dimensional Lebesgue measure.

\subsubsection{Underidentification}

\begin{definition}
If $r<p$ (so $m\leq 0$) then the system is called underidentified. \ 
\end{definition}

We split $\beta =\left( \beta _{1},...,\beta _{p}\right) ^{\prime }$ as $%
\beta _{\lbrack 1]}=\beta _{\mathcal{J}}$, $\beta _{\lbrack 2]}=\beta _{%
\mathcal{J}^{c}}$, where $\mathcal{J\cup J}^{c}=\left\{ 1,2,...,p\right\} $, 
$\dim (\mathcal{J})=p-r$ and $\dim (\mathcal{J}^{c})=r$, and build $\lambda
=(\theta _{1},...,\theta _{J-1},\beta _{\lbrack 1]}^{\prime })^{\prime }$, $%
\phi =\beta _{\lbrack 2]}$. Hence $\lambda $ augments $\theta $ with $p-r$
elements from $\beta $. Assumption A holds if $\mathcal{J}$ can be found
such that $\beta _{\lbrack 2]}=t(\theta _{1},...,\theta _{J-1},\beta
_{\lbrack 1]})$.

\begin{example}
Consider instrumental variables problem $g(s,\beta )=s^{(3)}\left\{
s^{(1)}-\beta ^{\prime }s^{(2)}\right\} $, $\dim (s^{(2)})=p$, $\dim
(s^{(3)})=r$. If $p>r$ then split $\beta =\left( \beta _{\lbrack 1]}^{\prime
},\beta _{\lbrack 2]}^{\prime }\right) ^{\prime }$, where $\dim (\beta
_{\lbrack 1]})=r-p$ and $\dim (\beta _{\lbrack 2]})=r$. Write $s_{j}=\left(
s_{j,[1]}^{\prime },s_{j,[2]}^{\prime }\right) ^{\prime }$, then 
\begin{equation*}
\sum_{j=1}^{J}\theta _{j}s_{j}^{(3)}\left\{ \left( s_{j}^{(1)}-\beta
_{\lbrack 1]}^{\prime }s_{j,[1]}^{(2)}\right) -\beta _{\lbrack 2]}^{\prime
}s_{j,[2]}^{(2)}\right\} =0_{r}.
\end{equation*}%
Knowledge of $\beta _{\lbrack 1]}$ puts us back to the just identified, so
Assumption A holds under weak assumptions and so $p(\theta ,\beta _{\lbrack
2]})$ can be computed using the area formula.
\end{example}

\subsubsection{Overidentification \ }

\begin{definition}
If $r>p$ so $m\geq 1$ (e.g. $r=2$,$p=1$, $m=2$) then the system is called
overidentified.\ \ 
\end{definition}

We split $\theta =\left( \theta _{1},...,\theta _{J-1}\right) ^{\prime }$ as 
$\theta _{\lbrack 1]}=\theta _{\mathcal{J}}$, $\theta _{\lbrack 2]}=\theta _{%
\mathcal{J}^{c}},$where $\mathcal{J\cup J}^{c}=\left\{ 1,2,...,J-1\right\} $%
, $\dim (\mathcal{J})=J-m$ and $\dim (\mathcal{J}^{c})=m-1$, and build $%
\lambda =\theta _{\lbrack 1]}^{\prime }$, $\phi =\left( \theta _{\lbrack
2]}^{\prime },\beta ^{\prime }\right) ^{\prime }$. Hence $\lambda $ is a
subset of $\theta $ with $J-m$ elements, while $\phi $ contains all the
other probabilities and the entire $\beta $. Then Assumption A holds if we
can find a $\mathcal{J}$ such that $\left( \theta _{\lbrack 2]}^{\prime
},\beta ^{\prime }\right) ^{\prime }=t(\theta _{\lbrack 1]}^{\prime })$.

\begin{example}
Again consider $g(s,\beta )=s^{(3)}\left\{ s^{(1)}-\beta ^{\prime
}s^{(2)}\right\} $, $\dim (s^{(2)})=p$, $\dim (s^{(3)})=r$. If $p<r$ then
split $\theta =\left( \theta _{\lbrack 1]}^{\prime },\theta _{\lbrack
2]}^{\prime }\right) ^{\prime }$, where $\dim (\theta _{\lbrack 2]}^{\prime
},\beta ^{\prime })=r$, so there are $r$ moment conditions and $r$ unknowns.
Given $\theta _{\lbrack 1]}$, we can then solve for the extended set of
parameters $\left( \theta _{\lbrack 2]}^{\prime },\beta ^{\prime }\right) $,
where 
\begin{equation*}
\sum_{j=1}^{\dim (\theta _{\lbrack 1]})}\theta _{j}s_{j}^{(3)}\left\{
s_{j}^{(1)}-\beta ^{\prime }s_{j}^{(2)}\right\} +\sum_{j=\dim (\theta
_{\lbrack 1]})+1}^{J}\theta _{j}s_{j}^{(3)}\left\{ s_{j}^{(1)}-\beta
^{\prime }s_{j}^{(2)}\right\} =0_{r}.
\end{equation*}%
This is typically exactly identified, but non-linear due to the $\theta
_{j}\beta $ terms for $j=\dim (\theta _{\lbrack 1]})+1,...,J$. \ 
\end{example}

\end{document}